\def\PI{Dessauges-Zavadsky et~al. 2003}
\def\PII{P\'eroux et~al. 2003b}
\def\zem{$z_{\rm em}$}
\def\zabs{$z_{\rm abs}$}

\def\kms{km~s$^{-1}$}

\def\cm2{cm$^{-2}$}
\def\loghi{\mbox{$\log N_{\rm HI}$}}
\def\nhi{\mbox{$N_{\rm HI}$}}
\def\hi{H~{\sc i}}

\def\e{et~al.}

\def\c2{C~{\sc ii}}
\def\c4{C~{\sc iv}}
\def\fe2{Fe~{\sc ii}}
\def\fe3{Fe~{\sc iii}}
\def\mg1{Mg~{\sc i}}
\def\mg2{Mg~{\sc ii}}
\def\si2{Si~{\sc ii}}
\def\si4{Si~{\sc iv}}
\def\al2{Al~{\sc ii}}
\def\al3{Al~{\sc iii}}
\def\o1{O~{\sc i}}
\def\o6{O~{\sc vi}}
\def\n1{N~{\sc i}}
\def\h1{H~{\sc i}}

\def\approxlt{\mathrel{\spose{\lower 3pt\hbox{$\sim$}}
        \raise 2.0pt\hbox{$<$}}}
\def\approxgt{\mathrel{\spose{\lower 3pt\hbox{$\sim$}}
        \raise 2.0pt\hbox{$>$}}}

\documentstyle[psfig,rotating]{mn}

\input{psfig.sty}

\newif\ifAMStwofonts

\title[Total Gas Mass $\Omega_{\rm HI+HeII}$ at $z>2$]{A Homogeneous Sample of Sub-DLAs III: \hspace{4cm} Total Gas Mass $\Omega_{\rm HI+HeII}$ at $z>2$\thanks{Based on observations collected during programme ESO 69.A-0613, ESO 71.A-0114 and ESO 73.A-0653 at the European Southern Observatory with UVES on the 8.2 m KUEYEN telescope operated at the Paranal Observatory, Chile.}}

\author[C\'eline P\'eroux et al.]
       {C\'eline P\'eroux$^{1}$\thanks{e-mail: cperoux@eso.org},
       Miroslava Dessauges-Zavadsky$^{2}$, Sandro D'Odorico$^{1}$,
\newauthor
Tae Sun Kim$^{3}$ \& Richard G. McMahon$^{3}$\\
$1$ European Southern Observatory, Karl-Schwarzchild-Str. 2, 85748 Garching-bei-M\"unchen, Germany.\\
$2$ Observatoire de Gen\`eve, 1290 Sauverny, Switzerland.\\
$3$ Institute of Astronomy, Madingley Road, Cambridge CB3 0HA, UK.\\
}

\date{}

\pagerange{\pageref{firstpage}--\pageref{lastpage}}

\begin{document}

\maketitle

\label{firstpage}

\begin{abstract}
Absorbers seen in the spectrum of background quasars are a unique tool
to select HI-rich galaxies at all redshifts. In turns, these allow to
determine the cosmological evolution of the \hi\ gas, $\Omega_{\rm
HI+HeII}$, a possible indicator of gas consumption as star formation
proceeds. The Damped Lyman-$\alpha$ systems (DLAs with \nhi $\ge$
10$^{20.3}$ cm$^{-2}$), in particular, are believed to contain a large
fraction of the \hi\ gas but there are also indications that lower
column density systems, named ``sub-Damped Lyman-$\alpha$'' systems
play a role at high-redshift. Here we present the discovery of
high-redshift sub-DLAs based on 17 $z>4$ quasar spectra observed with
the Ultraviolet-Visual Echelle Spectrograph (UVES) on VLT. This sample
is composed of 21 new sub-DLAs which, together with another 10 systems
from previous ESO archive studies, make up a homogeneous sample. The
redshift evolution of the number density of several classes of
absorbers is derived and shows that all systems seem to be evolving in
the redshift range from $z=5$ to $z\sim3$. This is further used to
estimate the redshift evolution of the characteristic radius of these
classes of absorbers assuming a Holmberg relation and one unique
underlying parent population. DLAs are found to have $R_*\sim20
h_{100}^{-1}$ kpc, while sub-DLAs have $R_*\sim40 h_{100}^{-1}$
kpc. The redshift evolution of the column density distribution,
f(N,z), down to \nhi = 10$^{19}$ cm$^{-2}$ is also presented. A
departure from a power law due to a flattening of f(N,z) in the
sub-DLA regime is present in the data. f(N,z) is further used to
determine the \hi\ gas mass contained in sub-DLAs at z$>$2. The
complete sample shows that sub-DLAs are important at all redshifts
from $z=5$ to $z=2$. Finally, the possibility that sub-DLAs are less
affected by the effects of dust obscuration than classical DLAs are
discussed.
\end{abstract}


\section{Introduction}

Tracing the rate at which stars form over cosmological scales still
remains a challenging observational task. An indirect way to probe the
assembly of galaxies is to probe the rate at which they convert their
gas into stars. The neutral \hi\ mass in particular can be estimated
from observations of absorbers seen in the spectrum of background
quasars. Unlike other high-redshift galaxies (such as Lyman Break
Galaxies, Steidel \e\ 2003), these objects are selected solely on
their \hi\ cross-sections regardless of their intrinsic luminosities
or star formation rates. The quasar absorption systems are divided
into several classes according to the number of atoms along the
observed line of sight: the Ly$\alpha$ forest have
\hi\ column densities ranging from $\simeq 10^{12}$ to $1.6 \times
10^{17}$ atoms cm$^{-2}$, the Lyman-limit systems (LLSs) with N(HI) $>
1.6 \times 10^{17}$ atoms cm$^{-2}$ and the damped Ly$\alpha$ systems
(DLAs) with N(HI) $> 2
\times 10^{20}$ atoms cm$^{-2}$.
These latter systems are believed to be the major contributors to the
neutral gas in the Universe at high redshifts. They thus can be used
to measure the redshift evolution of $\Omega_{\rm HI+HeII}$, the total
amount of neutral gas expressed as a fraction of today's critical
density (Lanzetta \e\ 1991; Wolfe \e\ 1995; Storrie-Lombardi et
al. 1996a; Storrie-Lombardi \& Wolfe 2000). We have recently suggested
(P\'eroux, et al. 2003a) that at $z>3.5$, some fraction of the \hi\
lies in systems below the traditional DLA definition, in ``sub-Damped
Lyman-$\alpha$ Systems (sub-DLAs)'' with $19$ $<$ log N(HI) $<$ 20.3
cm$^{-2}$. The name arises from the fact that this column density
range lies on the linear part of the curve of growth which links the
equivalent width of an absorber to the number of atoms along the line
of sight. The present paper reviews these predictions based on direct
observations of a sample of sub-DLAs.

In addition, quasar absorbers are a direct probe of element abundances
over $ > 90 \%$ of the age of the Universe. The cosmological evolution
of the HI column density weighted metallicity (e.g., Kulkarni \& Fall
2002) shows surprising results: contrary to virtually all chemical
models (e.g., Malaney \& Chaboyer 1996; Pei, Fall \& Hauser 1999), the
most recent observations indicate mild evolution with redshift (Khare,
et~al. 2004; Kulkarni, et~al. 2005). Therefore, the results to date
might give a biased or incomplete view of the global galactic chemical
evolution. The ultimate goal of this project is to undertake detailed
abundance and dynamical studies of an appropriate sample of sub-DLAs
in order to estimate the evolution of their mean abundance with
redshift.

In a first step towards this goal, we took advantage of the ESO VLT
archive to build a sample of sub-DLAs by analysing UVES archival
echelle QSO spectra. This represents a sample of 35 QSOs, 22 of which
were unbiased for our study. This study led to the discovery of 12
sub-DLAs (Dessauges-Zavadsky et~al. 2003). Their chemical abundances
were derived using Voigt profile fitting and photoionization models
from the CLOUDY software package in order to determine the ionization
correction. We find that the correction is negligible in systems with
$N(HI) > 3.2 \times 10^{19}$ and lower than 0.3 dex for most element
in systems with $10^{19} <$ N(HI) $< 3.2
\times 10^{19}$ atoms cm$^{-2}$. These systems were used to
observationally determine the shape of the column density
distribution, $f(N,z)$, down to N(HI) $=10^{19}$ cm$^{-2}$ (P\'eroux
et~al. 2003b), although the lack of high-redshift systems prevented from
measuring the $f(N,z)$ redshift evolution. The abundances observed in
this sample of sub-DLAs were further used to determine the global
metallicity of \hi\ gas in both DLAs and sub-DLAs.

Here we present a new sample of 17 z$>$4 quasar lines of sight
observed at high-resolution with the Ultraviolet-Visual Echelle
Spectrograph (UVES) on VLT. We use these newly acquired data to search
for and study the statistical properties of high-redshift sub-DLAs. In
the second section, the observations and data reduction process are
detailed together with a description of each object. The third
section presents the statistical characteristics of the survey and the
newly built sub-DLA sample. Finally, our current state of knowledge of
this class of absorbers together with implications for the
cosmological evolution of \hi\ gas mass are discussed in section 4.

\section{The $z>2$ Sub-DLA Sample}

\subsection{Observations and Data Reduction}

\begin{table*}
\begin{center}
\caption{Journal of observations of our observing programmes to find and 
analyse new high-redshift sub-DLAs in 17 $z>4$ quasars.\label{t:JoO}}
\begin{tabular}{llclllll}
\hline
Quasar &Alternative name	&$z_{\rm em}$ &R mag &Obs Date &UVES settings &Exp. Time (sec)	&Ref	\\
\hline
BR J0006$-$6208      &...		    &4.455   &19.25	&27 Sept 2002	&580           &7200      &1	\\      
PSS J0118$+$0320$^{red}$ &...		    &4.230   &18.66	&28 Sept 2002	&560           &3600	  &6	\\	
...                  &...		    &...     &...       &01 Sept 2003	&390+580       &5400      &...	\\
...                  &...		    &...     &...       &01 Sept 2003	&580	       &5400      &...	\\
...                  &...		    &...     &...       &01 Sept 2003	&860	       &5400$\times$2&...	\\
PSS J0121$+$0347$^{red*}$ &...		    &4.127   &18.33	&28 Sept 2002	&540           &3600      &5	\\      
...                  &...		    &...     &...       &02 Sept 2003	&470+800       &3600$\times$2&...	\\
SDSS J0124$+$0044$^{red}$&...		    &3.840   &17.75	&27 Sept 2002	&520           &3600      &4	\\      
...                  &...		    &...     &...       &02 Sept 2003	&520	       &3600	  &...	\\
...                  &...		    &...     &...       &29/04 Aug/Sept 2004	&580   &3000$\times$4	  &...	\\
PSS J0133$+$0400     &...		    &4.154   &18.32	&27 Sept 2002	&520           &3600	  &1,7	\\	
BRI J0137$-$4224$^{red}$ &BRI B0135$-$4239	    &3.970   &18.77	&27 Sept 2002	&520           &3600	  &3	\\	
...                  &...		    &...     &...       &03 Sept 2003	&520	       &5000      &...	\\
...                  &...		    &...     &...       &03 Sept 2003	&470+800       &3600+4500 &...	\\
PSS J0209$+$0517     &...		    &4.174   &17.76	&28 Sept 2002	&390+580       &7200	  &1,7	\\         
BRI J0244$-$0134     &BRI B0241$-$0146	    &4.053   &18.18	&27 Sept 2002	&520           &3600	  &2	\\	
BR J0311$-$1722      &...		    &4.039   &17.73	&28 Sept 2002	&540           &3600      &1	\\      
BR J0334$-$1612$^{red}$  &...		    &4.363   &17.86	&28 Sept 2002	&560           &3600	  &1	\\	
...                  &...		    &...     &...       &02 Sept 2003	&580	       &3600$\times$2&...	\\
...                  &...		    &...     &...       &02 Sept 2003	&860	       &5066	  &...	\\
BR J0419$-$5716      &...		    &4.461   &17.78	&28 Sept 2002	&560           &1846      &1	\\      
BR J2017$-$4019$^{BAL}$  &BRLCO B2013$-$4028    &4.131   &18.60	&03 Sept 2003	&540	       &3600	  &1	\\         
PSS J2154$+$0335     &...		    &4.363   &19.05	&28 Sept 2002	&560           &3600	  &1	\\	
BR J2215$-$1611$^{red}$  &BR B2212$-$1626	    &3.990   &18.10	&28 Sept 2002	&540           &3600	  &2	\\	
...                  &...		    &...     &...       &03 Sept 2003	&540	       &3600+1215 &...	\\
...                  &...		    &...     &...       &03 Sept 2003	&470+800       &3600$\times$2&...	\\
BR J2216$-$6714$^{red}$  &...		    &4.469   &18.57	&28 Sept 2002	&560           &3600	  &1	\\	
...                  &...		    &...     &...       &01 Sept 2003	&560	       &3600$\times$2&...	\\
...                  &...		    &...     &...       &01 Sept 2003	&580	       &5400      &...	\\
...                  &...		    &...     &...       &02 Sept 2003	&860	       &5400$\times$2&...	\\
BR J2239$-$0552      &BR B2237$-$0607	    &4.558   &18.30	&27 Sept 2002	&580           &3600	  &2	\\	
BR J2349$-$3712      &BRLCO B2346$-$3729    &4.208   &18.70	&03 Sept 2003	&540	       &3600	  &1	\\         
\hline
\end{tabular}
\end{center}
\vspace{0.2cm}
\begin{minipage}{140mm}
{\bf $^{BAL}$:} affected by Broad Absorption Lines features.\\
{\bf $^{red}$:} spectrum with complete wavelength coverage including to the red of the Ly-$\alpha$ emission line.\\
{\bf $^{red*}$:} attempt to get a spectrum with complete wavelength coverage impaired by bad weather conditions (strong northern wind).\\
{\bf References} -- (1) P\'eroux \e\ 2001; (2) Storrie-Lombardi \e\ 1996a; (3) Storrie-Lombardi \e\ 2001; (4) Schneider \e\ 2002; (5) Stern \e\ 2000; (6) from the Palomar Sky Survey, see http://www.astro.caltech.edu/$\sim$george/z4.qsos; (7) Prochaska et al. 2003.\\
\end{minipage}
\end{table*}

In order to complement the previous study and analyse sub-DLAs at $z >
3$, we started in September 2002 to build a UVES sample of high
redshift QSOs never observed before at high resolution. This snapshot
is carefully designed to determine the redshift evolution of the
statistical properties of sub-DLAs (number density and column density
distribution). The second step of the observing programme undertook a
year later, in September 2003, aims at concentrating on the 7 most
promising targets to enable detailed metallicity and dynamical studies
at high-redshifts. An extra set of data was obtained on August
29$^{\rm th}$ 2004, September 2$^{\rm nd}$ and 4$^{\rm th}$ 2004 in
service mode in the framework of a parallel observing programme (ESO
73.A-0653; PI: N. Bouch\'e). Table~\ref{t:JoO} gives the journal of
the observations.

The data reduction of the observations from period 69 and 71 is done
using the version of the UVES pipeline within the MIDAS environment
available at the time (version: uves/2.0.0 flmidas/1.0.0). The newest
data (period 73) are reduced with the most recent version of the pipeline
to accommodate for new format of the raw fits file (version: uves/2.1.0
flmidas/1.1.0). The raw wavelength frames are inspected closely to
optimise the number of order to be extracted for each CCD. Master bias
and flat images are constructed using calibration frames taken the
closest in time to the science frames. Lamps are taken on various
occasions through the observing nights to be able to make accurate
wavelength calibrations. In most cases, the science frames and
associated stars are extracted with the ``optimal''
option. Nevertheless, on a few frames from the second observing run
(September 2003), two of the settings (540 and 800) show a periodic
feature of the order of a few Angstr\"oms. These features are not
present when these few frames are extracted using the ``average''
option followed by a cosmic filtering.

For the 7 objects with more than one setting, the resulting spectra
are combined by using the signal-to-noise of the spectrum as a
weight. The final spectra have resolution of 7 km s$^{-1}$.  In some
case, we notice that the part of the spectra fully absorbed do not
actually reach the zero flux level. This problem is known to occur
when there are not enough photons in the sky area. Indeed, the optimal
extraction is done to fit a Gau\ss ian profile and the baseline of the
Gau\ss ian is the sky level. If the seeing during the observations is
poor and the exposure short, it is possible that at blue wavelengths,
only a few photons fall on the sky pixel. This leads to a high photon
noise and might produce the under-subtraction of the sky level. The
unsaturated absorption lines are unaffected by this zero-level
problem. In the case of saturated lines, we correct the spectrum by
subtracting a few percent from the continuum level. The spectra are
then normalised using a spline function to join the part of the
continuum free from absorption lines.

\subsection{Absorbers Identification and Column Density Measurements}

\begin{table*}
\begin{center}
\caption{High-redshift quasar absorber sample composed of 21 sub-DLAs and 7 DLAs\label{t:HI}.}
\begin{tabular}{lccclll}
\hline
Quasar &\zem &\zabs &\loghi &Ly series &metals &Note\\
\hline
BR J0006$-$6208    	&4.455  &3.202	&20.80$\pm$0.10 &Ly1       &no metals over available coverage  &DLA  \\
...                	&...    &4.145	&19.37$\pm$0.15 &Ly3       &no metals over available coverage  &...  \\
PSS J0118$+$0320$^{red}$  &4.230  &4.128  &20.02$\pm$0.15 &Ly5     &FeII, SiII, OI, CII, SiIV, CIV   &...  \\
PSS J0121$+$0347$^{red*}$ &4.127  &2.976  &19.53$\pm$0.10 &Ly1	   &OI, CIV, FeII, AlII              &...  \\
SDSS J0124$+$0044$^{red}$ &3.840  &2.988  &19.18$\pm$0.10 &Ly1	   &SiII, CIV, FeII, AlII            &...  \\
...                     &...    &3.078  &20.21$\pm$0.10 &Ly1	   &CII, SiII, OI, SiIV, CIV, FeII   &...  \\
PSS J0133$+$0400        &4.154  &3.139	&19.01$\pm$0.10	&Ly1	   &NiII, CII                        &...  \\
...                     &...	&3.692	&20.68$\pm$0.15	&Ly2	   &no metals over available coverage  &DLA  \\
...                     &...	&3.773  &20.42$\pm$0.10	&Ly2	   &no metals over available coverage  &DLA  \\
...                     &...    &3.995	&19.94$\pm$0.15	&Ly4	   &no metals over available coverage  &blended with the following system  \\
...                     &...    &3.999	&19.16$\pm$0.15	&Ly4	   &no metals over available coverage  &blended with the previous system  \\
...                     &...    &4.021	&19.09$\pm$0.15	&Ly4	   &FeII	                     &...  \\
BRI J0137$-$4224$^{red}$ &3.970  &3.101	&19.81$\pm$0.10	&Ly4	   &SiII, CIV, FeII, AlII            &...  \\
...                     &...    &3.665	&19.11$\pm$0.10	&Ly11	   &SiII, OI, CII, SiIV              &...  \\
PSS J0209$+$0517        &4.174  &3.666	&20.47$\pm$0.10	&Ly2       &SiIV                             &DLA  \\
...                     &...    &3.707  &19.24$\pm$0.10 &Ly2       &OI, SiIV                         &...  \\
...                     &...    &3.863	&20.43$\pm$0.15 &Ly3       &SiII, OI, CII                    &DLA  \\
BRI J0244$-$0134        &4.053  &...    &...            &...	   &...			             &...  \\
BR J0311$-$1722         &4.039  &3.734	&19.48$\pm$0.10	&Ly7	   &SiII, OI, CII                    &...  \\
BR J0334$-$1612$^{red}$ &4.363  &3.557	&21.12$\pm$0.15	&Ly1	   &SiII, CIV, FeII, AlII, ZnII      &DLA  \\
BR J0419$-$5716         &4.461  &3.063	&19.17$\pm$0.10	&Ly1	   &OI	                             &...  \\
BR J2017$-$4019$^{BAL}$ &4.131  &...    &...            &...	   &...			             &...  \\
PSS J2154$+$0335        &4.363  &3.177  &19.23$\pm$0.15 &Ly1       &no metals over available coverage  &...  \\
BR J2215$-$1611$^{red}$ &3.990  &3.656  &19.01$\pm$0.15 &Ly3       &NiII, CII, ZnII 	             &blended with the following system  \\
...                     &...    &3.662  &20.05$\pm$0.15 &Ly3       &OI, CII, SiIV, SiII, CIV, FeII, AlII &blended with the previous system  \\
BR J2216$-$6714$^{red}$ &4.469  &3.368  &19.80$\pm$0.10 &Ly1       &CII, CIV, AlII                   &...  \\
BR J2239$-$0552         &4.558  &4.079  &20.55$\pm$0.10 &Ly3       &SiII, OI, CII                    &DLA  \\
BR J2349$-$3712         &4.208  &3.581	&19.12$\pm$0.10	&Ly2	   &no metals over available coverage  &...  \\     
...                     &...    &3.690	&19.79$\pm$0.15	&Ly2	   &no metals over available coverage  &blended with the following system  \\ 
...                     &...    &3.696	&19.78$\pm$0.10	&Ly2	   &no metals over available coverage  &blended with the previous system  \\ 
\hline			
\end{tabular}
\end{center}
\vspace{0.2cm}
\begin{minipage}{140mm}
{\bf $^{BAL}$:} affected by Broad Absorption Lines features.\\
{\bf $^{red}$:} spectrum with complete wavelength coverage including to the red of the Ly-$\alpha$ emission line.\\
{\bf $^{red*}$:} attempt to get a spectrum with complete wavelength coverage impaired by bad weather conditions (strong northern wind).\\
\end{minipage}
\end{table*}

The methodology to search for sub-DLAs in the spectra described above
very much follows the one from our previous studies (\PI). We first
use an automated detection algorithm supplemented further by
eye-balling searches independently undertook by three of us (C. P.,
M. D.-Z. and T.-S. K.). Although the forest of the quasars at these
redshifts is considerably absorbed by foreground structures, the
presence of damping wings down to the Ly2 and occasionally to higher
lines of the series provides an unambiguous signature of damped
absorbers. We also report the detection of metal lines at the
associated redshifts, although this is not used as a criterion for
sub-DLA selection. Table~\ref{t:HI} summarises these information
together with the redshifts, \hi\ column densities and the references
of the 7 DLAs and 21 sub-DLAs found in our new high-redshift quasar
sample. The latter systems span a range of \hi\ column densities from
$10^{19.01}$ to $10^{20.21}$ cm$^{-2}$ with \zabs=2.976 to 4.145.

The \hi\ column density are determined by fitting a Voigt profile to
the absorption line. The fits were performed using the $\chi^2$
minimisation routine {\tt fitlyman} in {\tt MIDAS} (Fontana \&
Ballester 1995). The Doppler parameter $b$-value was usually fixed at
20 km~s$^{-1}$ or left as a free parameter since in that high column
density regime, the sub-DLA systems \nhi\ are independent of the
$b$-value. The fit is performed using the higher members of the Lyman
series where these are available. The typical resulting error bar in
\loghi\ measurements is 0.10 and never exceeds 0.15, but does not
include error in the continuum placement which we expect should not
exceed 10\%.

\subsection{Notes on Individual Objects}

In this section, we provide details on each of the identified
absorbers. Whenever the spectral coverage is available, we search for
metals associated with the Ly lines among the 20 transitions most
frequently detected in high column density quasar absorbers.

\begin{enumerate}

\item{BR J0006$-$6208 (\zem=4.455): P\'eroux \e\ (2001) have reported four
candidate absorbers at \zabs$=$2.97, 3.20, 3.78 and 4.14 in the medium
resolution spectrum of this quasar. The blue end of the spectrum
presented here is of fairly poor quality and therefore the minimum
redshift above which quasar absorbers are searched for is set to
z$_{\rm min}$=3.083, so the lowest DLA cannot be checked upon. We
confirm the presence of the \zabs=3.202 system and measure
\loghi=20.80$\pm$0.10 in agreement with medium resolution
estimates. The 3${rd}$ system falls in the UVES setting gap. The last
absorber is confirmed as a sub-DLA with \loghi=19.37$\pm$0.15. No
associated metal lines are detected over the limited wavelength
coverage of our spectrum. Figure~\ref{f:Q0006} shows the best Voigt
profile fitting solutions for these two systems.}

\begin{figure}
\begin{center}
\includegraphics[height=4cm, width=8cm, angle=0]{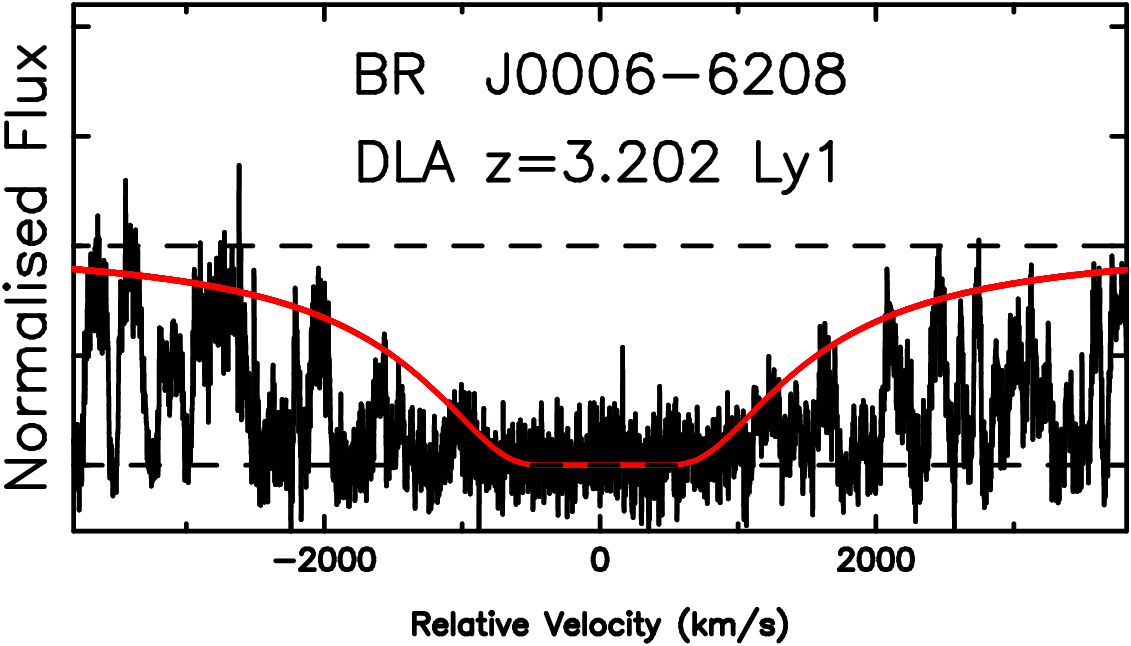}
\vspace{0.5cm}
\includegraphics[height=9cm, width=8cm, angle=0]{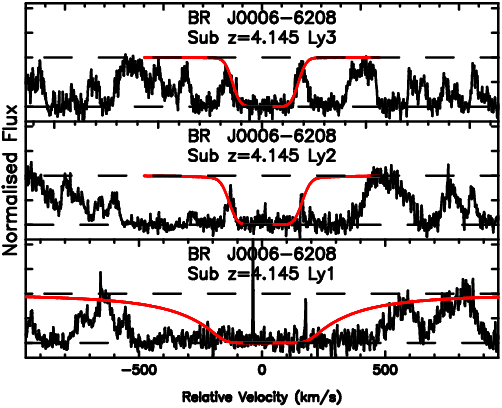}
\caption{The two absorbers detected towards BR J0006$-$6208 at 
\zabs=3.202 (\loghi=20.80$\pm$0.10) and \zabs=4.145 
(\loghi=19.37$\pm$0.15).\label{f:Q0006}} 
\end{center}
\end{figure}

\vspace{0.5cm}
\item{PSS J0118$+$0320 (\zem=4.230): This quasar was discovered as part 
of the Palomar Sky Survey and to our knowledge, there are no medium
resolution spectrum published. We discover an unambiguous sub-DLA at
\zabs=4.128 and measure \loghi=20.02$\pm$0.15 down to the Lyman
5. Several associated metals are also observed in the red part of the
spectrum. Figure~\ref{f:Q0118} shows our best fit of the \hi\ lines.}

\begin{figure}
\begin{center}
\includegraphics[height=11cm, width=8cm, angle=0]{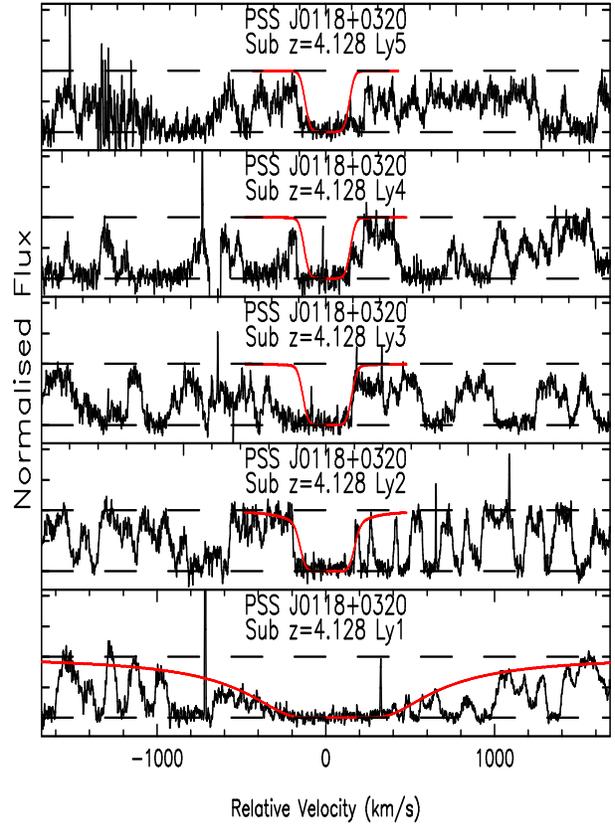}
\caption{The absorber detected towards PSS J0118$+$0320 st \zabs=4.128 (\loghi=20.02$\pm$0.15).\label{f:Q0118}} 
\end{center}
\end{figure}

\vspace{0.5cm}
\item{PSS J0121$+$0347 (\zem=4.127): No absorbers have been previously 
reported in this quasar also from the Palomar Sky Survey. We report
here for the first time a sub-DLA with \loghi=19.53$\pm$0.10 at
\zabs=2.976. Metal lines at that redshift are also observed in the red
part of the spectrum. Figure~\ref{f:Q0121} shows our best fit of the
\hi\ line.}

\begin{figure}
\begin{center}
\includegraphics[height=4cm, width=8cm, angle=0]{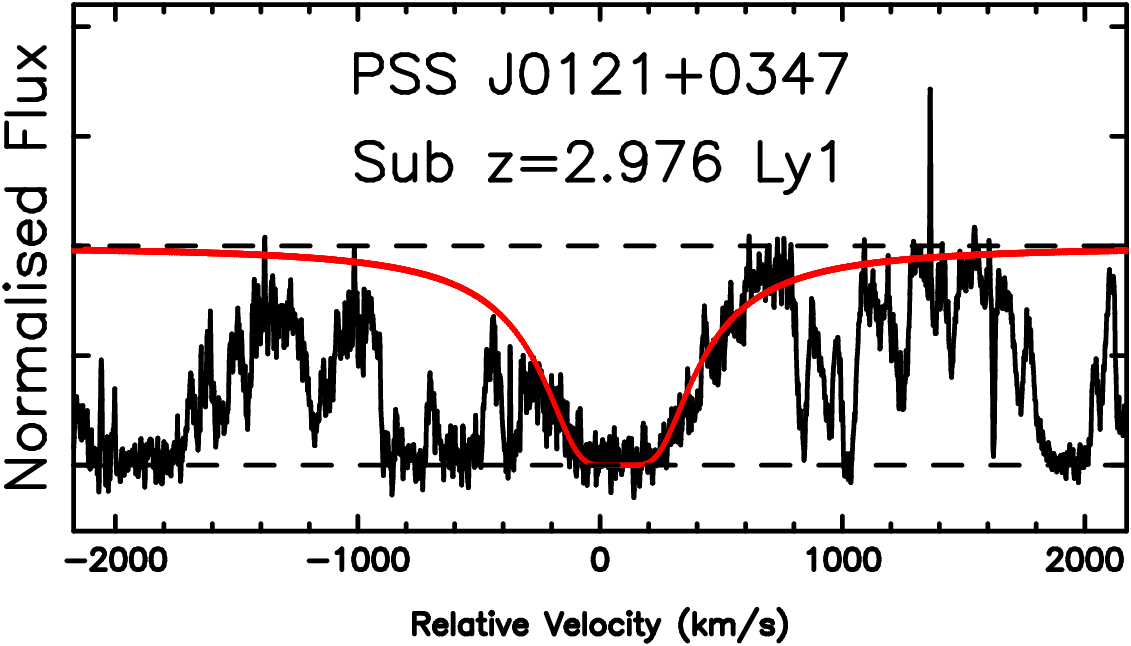}
\caption{The absorber detected towards PSS J0121$+$0347 at \zabs=2.976 (\loghi=19.53$\pm$0.10).\label{f:Q0121}} 
\end{center}
\end{figure}

\vspace{0.5cm}
\item{SDSS J0124$+$0044 (\zem=3.840): One absorber have been previously 
reported in this quasar from the Sloan Digital Sky Survey (Bouch\'e
\& Lowenthal, 2004). We report here for the first time the column density of the two sub-DLAs with \loghi=19.18$\pm$0.10 at
\zabs=2.988 and \loghi=20.21$\pm$0.10 at
\zabs=3.078. Metal lines at these redshifts are also observed in the red
part of the spectrum. Figure~\ref{f:Q0124} shows our best fit of the
\hi\ lines.}

\begin{figure}
\begin{center}
\includegraphics[height=4cm, width=8cm, angle=0]{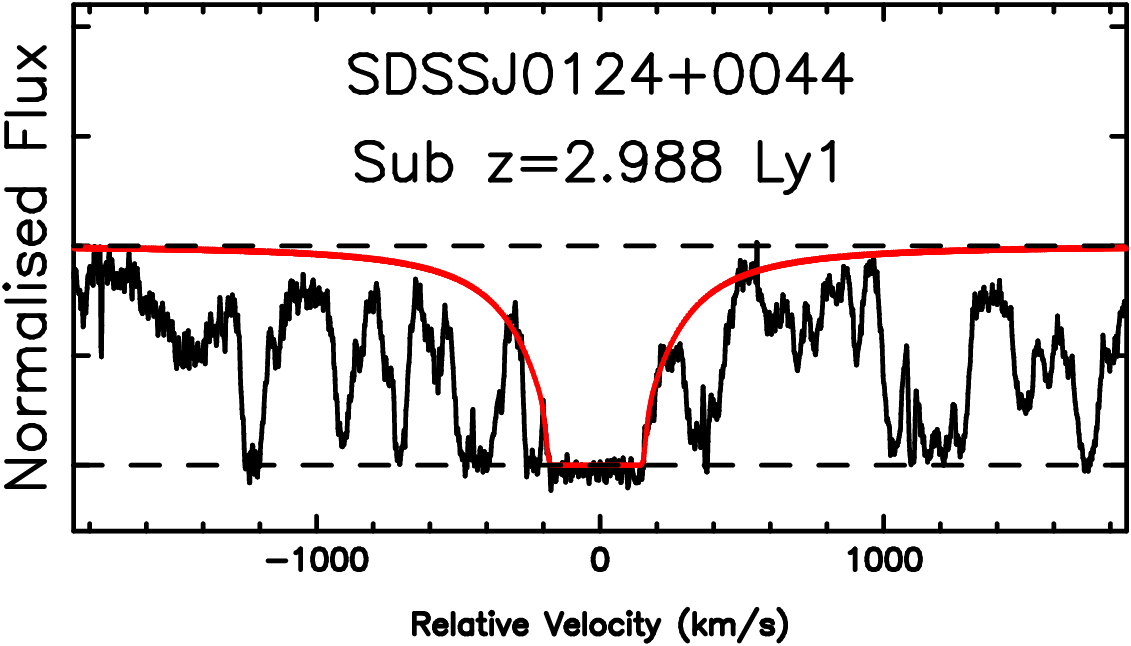}
\vspace{0.5cm}
\includegraphics[height=4cm, width=8cm, angle=0]{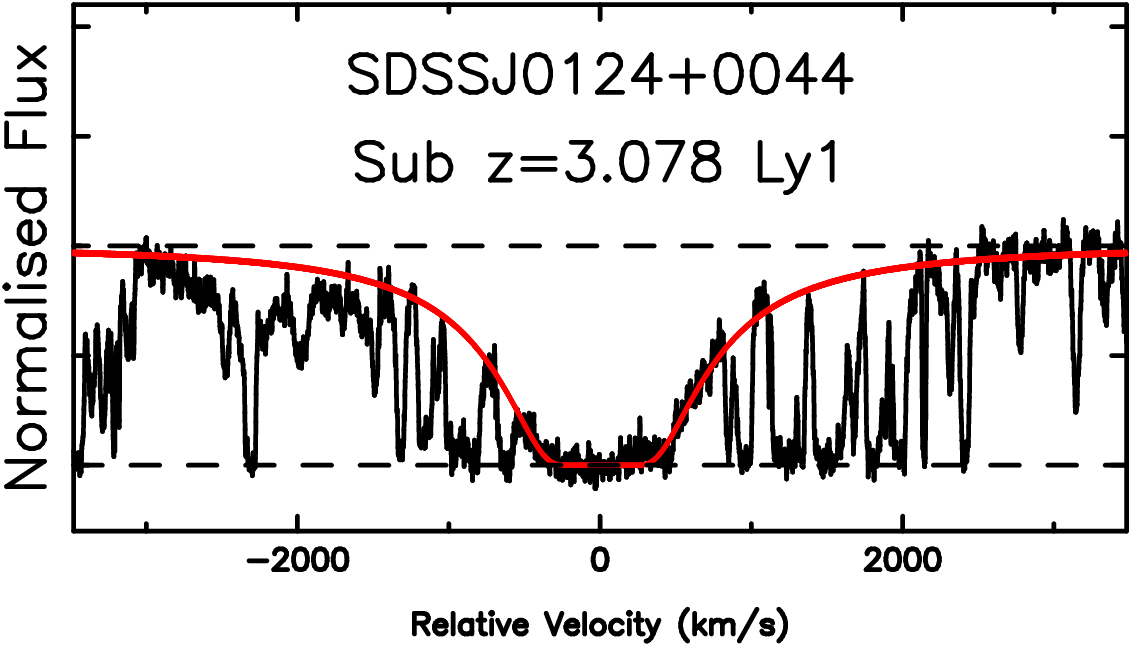}
\caption{The absorbers detected towards SDSS J0124$+$0044 at \zabs=2.988 (\loghi=19.18$\pm$0.10) and \zabs=3.078 (\loghi=20.21$\pm$0.10).
\label{f:Q0124}} 
\end{center}
\end{figure}

\vspace{0.5cm}
\item{PSS J0133$+$0400 (\zem=4.154): P\'eroux \e\ (2001) report two DLAs in 
this quasar which were further confirmed by ESI/Keck II observations
of Prochaska \e\ (2003). Here, we measure \loghi=20.68$\pm$0.15 at
\zabs=3.692 and \loghi=20.42$\pm$0.10 at \zabs=3.773 (Prochaska \e\
measure \loghi=20.70$_{\rm -0.15}^{\rm +0.10}$ and
\loghi=20.55$_{\rm -0.15}^{\rm +0.10}$ respectively). We also note 
that these two systems are close by each other: there are separated by
$\sim$ 5170 \kms with an additional \loghi$<$19.0 in between them (see
second panel of Figure~\ref{f:Q0133}). This group constitutes almost a
multiple DLA (Lopez \e\ 2001, Ellison \& Lopez 2001, Lopez
\& Ellison 2003). In addition, we report the discovery of a further 4
sub-DLAs along the same line of sight. We measure
\loghi=19.01$\pm$0.10 at \zabs=3.139, \loghi=19.94$\pm$0.15 at
\zabs=3.995, \loghi=19.16$\pm$0.15 at \zabs=3.999 and \loghi=19.09$\pm$0.15 
at \zabs=4.021. Again, two of these are very close by \zabs=3.995 and
3.999 corresponding to 250 \kms. There are fitted together using Ly2
and Ly4 since those provide better constrains than Ly1. The Lyman
series down to Ly4 is also available for the last 3 sub-DLAs. The
various Lyman series absorption lines along this rich line of sight
are displayed in Figure~\ref{f:Q0133}. Given the limited wavelength
coverage of our spectrum, only metals associated with the lowest and
the highest system are detected.}

\begin{figure*}
\begin{center}
\vspace{0.5cm}
\includegraphics[height=4cm, width=8cm, angle=0]{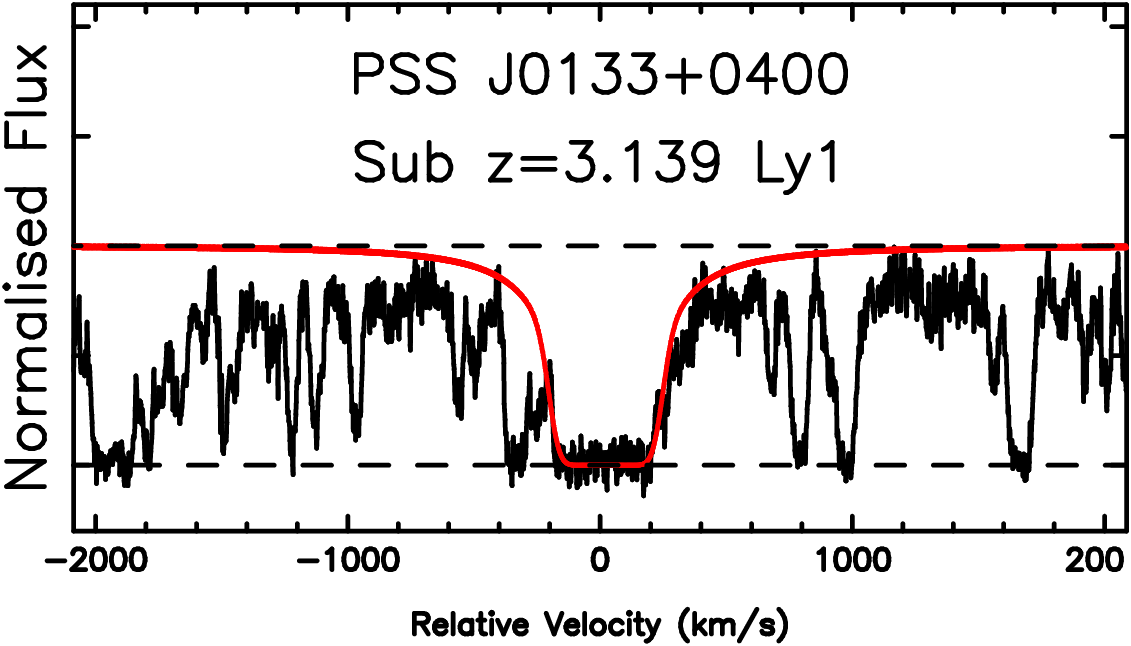}
\end{center}
\includegraphics[height=7cm, width=8cm, angle=0]{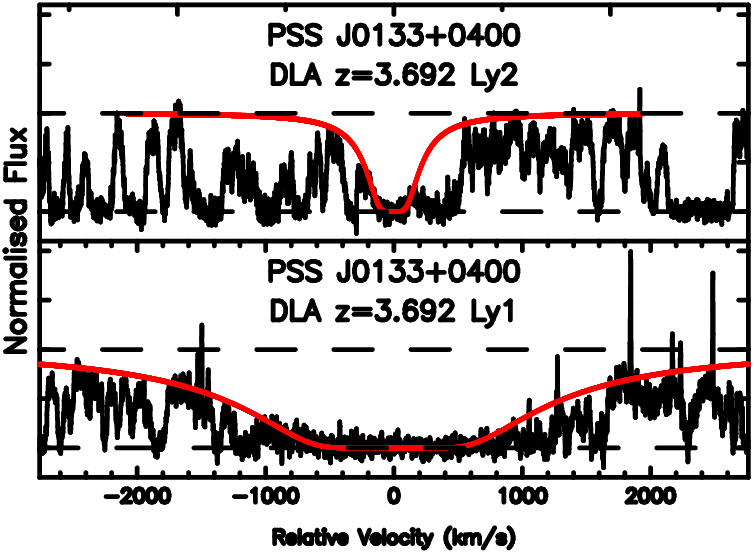}
\vspace{0.5cm}
\includegraphics[height=7cm, width=8cm, angle=0]{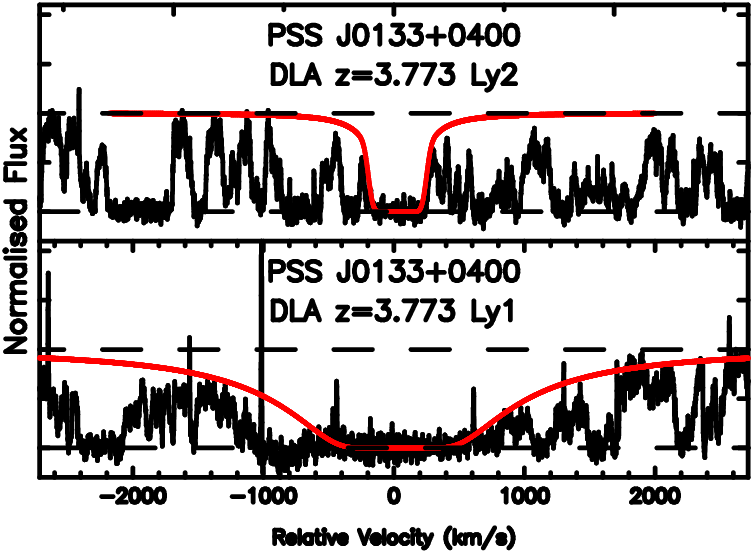}
\vspace{0.5cm}
\includegraphics[height=10cm, width=8cm, angle=0]{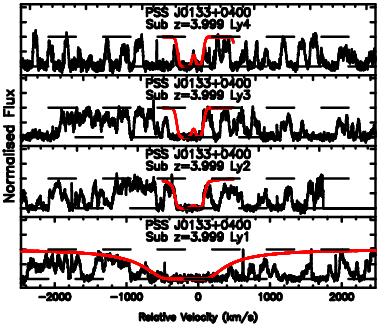}
\vspace{0.5cm}
\includegraphics[height=10cm, width=8cm, angle=0]{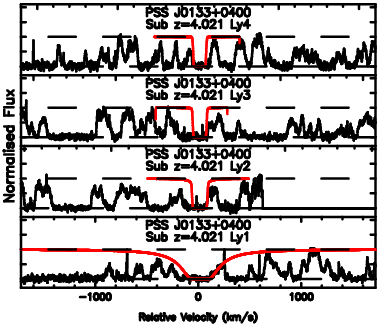}
\caption{The rich line of sight towards PSS J0133$+$0400 is composed of 
the systems with following redshifts: \zabs=3.139 (\loghi=19.01$\pm$0.10),
\zabs=3.692 (\loghi=20.68$\pm$0.15), \zabs=3.773 (\loghi=20.42$\pm$0.10), 
\zabs=3.995 (\loghi=19.94$\pm$0.15) together with  
\zabs=3.999 (\loghi=19.16$\pm$0.15) and finally \zabs=4.021 (\loghi=19.09$\pm$0.15).\label{f:Q0133}}
\end{figure*}

\vspace{0.5cm}
\item{BRI J0137$-$4224 (\zem=3.970): Storrie-Lombardi \e\ (2001) do 
not find any DLA in the spectrum of this quasar. We confirm that there
are none of the highest \hi\ column densities, but we find two
sub-DLAs. The first system is at \zabs=3.101 and has a \hi\ column
density
\loghi=19.81$\pm$0.10, while the second is at \zabs=3.665 and has
\loghi=19.11$\pm$0.10. Both these systems have metals associated with
them. The lyman series down to Ly11 is also visible for the second
system. Figure~\ref{f:Q0137} shows the best Voigt profile fitting
solutions for these two systems.}

\begin{figure}
\begin{center}
\includegraphics[height=4cm, width=8cm, angle=0]{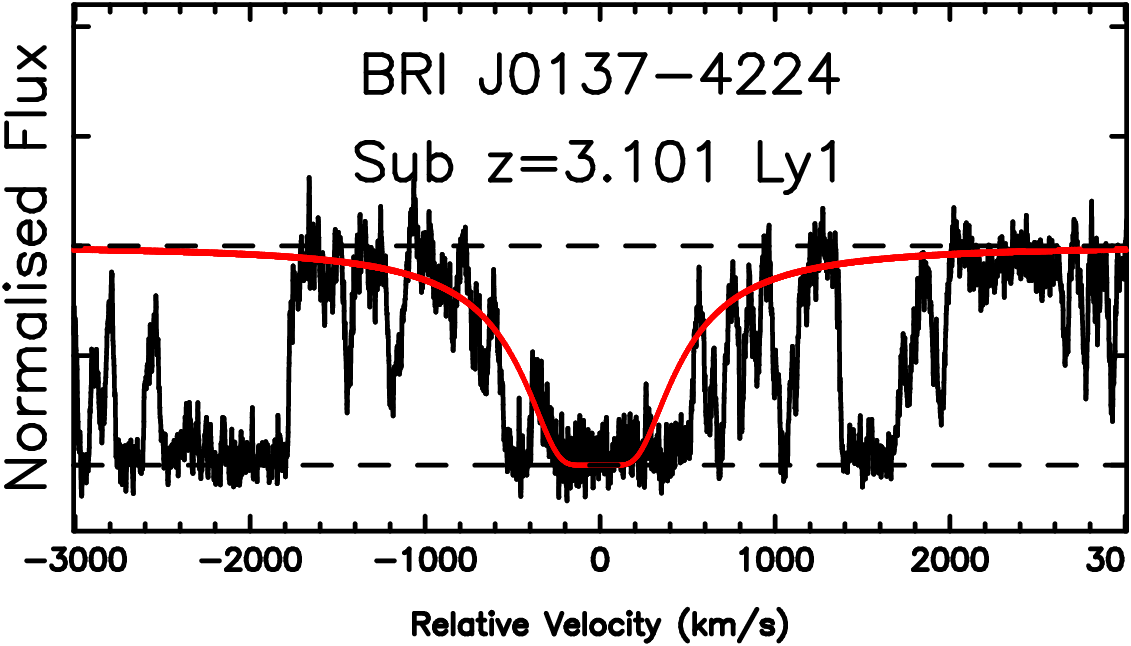}
\vspace{0.5cm}
\includegraphics[height=16.5cm, width=8cm, angle=0]{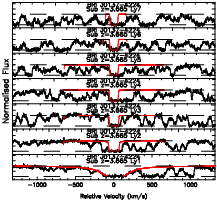}
\caption{The two absorbers detected towards BRI J0137$-$4224 at 
\zabs=3.101 (\loghi=19.81$\pm$0.10) and \zabs=3.665 (\loghi=19.11$\pm$0.10). The latter system is detected down to Ly11, but for display purpose only
the first 7 members of the series are shown.\label{f:Q0137} }
\end{center}
\end{figure}

\vspace{0.5cm}
\item{PSS J0209$+$0517 (\zem=4.174): P\'eroux \e\ (2001) report two DLAs in 
this quasar which were further confirmed by ESI/Keck II observations
of Prochaska \e\ (2003). Here, we measure \loghi=20.47$\pm$0.10 at
\zabs=3.666 and \loghi=20.45$\pm$0.15 at \zabs=3.863 (Prochaska \e\
measure \loghi=20.55$\pm$0.10 and \loghi=20.55$\pm$0.10
respectively). We also find a sub-DLA at \zabs=3.707. We measure
\loghi=19.24$\pm$0.10 for that system. All these absorbers have metals 
associated with them. Figure~\ref{f:Q0209} shows our best fit of these
\hi\ lines.}

\begin{figure}
\begin{center}
\includegraphics[height=6cm, width=8cm, angle=0]{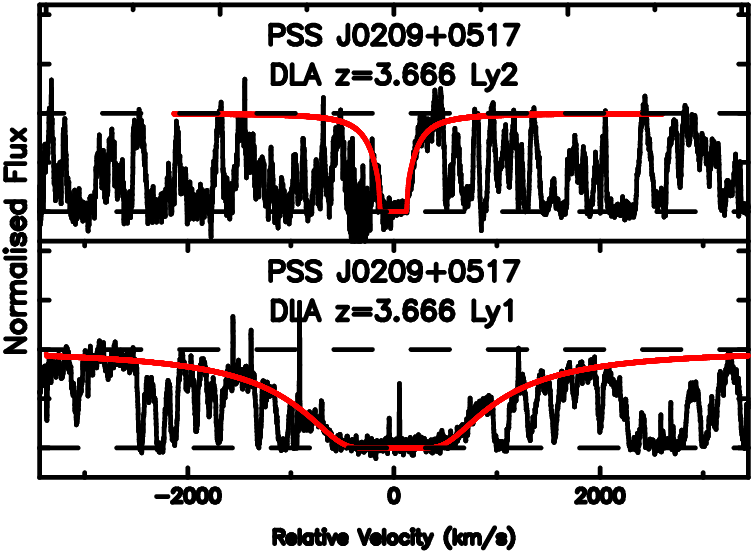}
\vspace{0.5cm}
\includegraphics[height=6cm, width=8cm, angle=0]{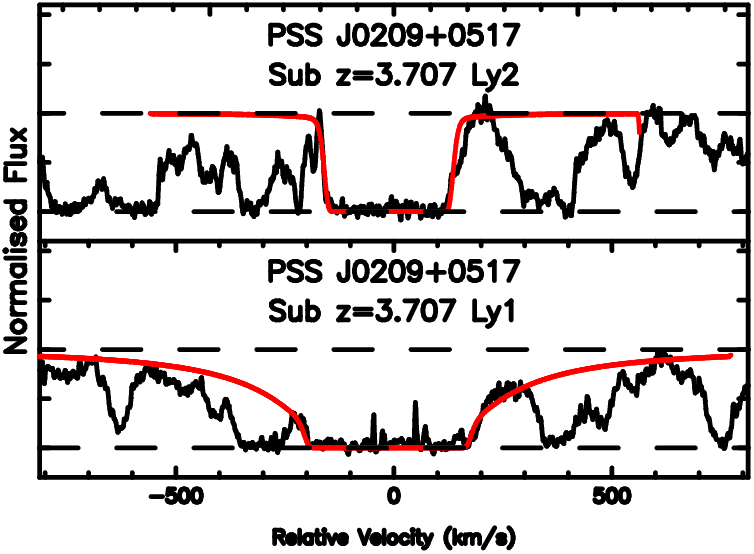}
\vspace{0.5cm}
\includegraphics[height=8cm, width=8cm, angle=0]{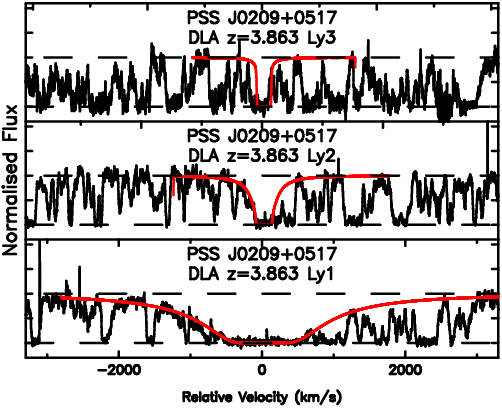}
\vspace{0.5cm}
\caption{The absorbers detected towards PSS J0209$+$0517 at 
\zabs=3.666 (\loghi=20.47$\pm$0.10), \zabs=3.707 (\loghi=20.55$\pm$0.10) 
and \zabs=3.863 (\loghi=20.45$\pm$0.15).\label{f:Q0209}}
\end{center}
\end{figure}

\vspace{0.5cm}
\item{BR J0244$-$0134 (\zem=4.053): This quasar was observed by 
Storrie-Lombardi \e\ (1996b) who did not report any DLA. The
high-resolution spectrum that we have acquired shows that no sub-DLA
is found either along this line of sight.}

\vspace{0.5cm}
\item{BR J0311$-$1722 (\zem=4.039): P\'eroux \e\ (2001) report an absorber 
with column density below the classical definition along this line of
sight from medium resolution spectroscopy. Here, we confirm that the
system has \loghi=19.48$\pm$0.10 at \zabs=3.734 observable down to Ly7
as shown in Figure~\ref{f:Q0311}. Most of the constraints in this system
comes from Ly3 and Ly5. The lower members of the series were therefore
merely used as a check for the solution. We note that one using the
Ly1 only to fit this system would mistakingly derive a much higher \hi\
($>$20.0). Metals are found at the same redshift.}

\begin{figure}
\begin{center}
\includegraphics[height=16.5cm, width=8cm, angle=0]{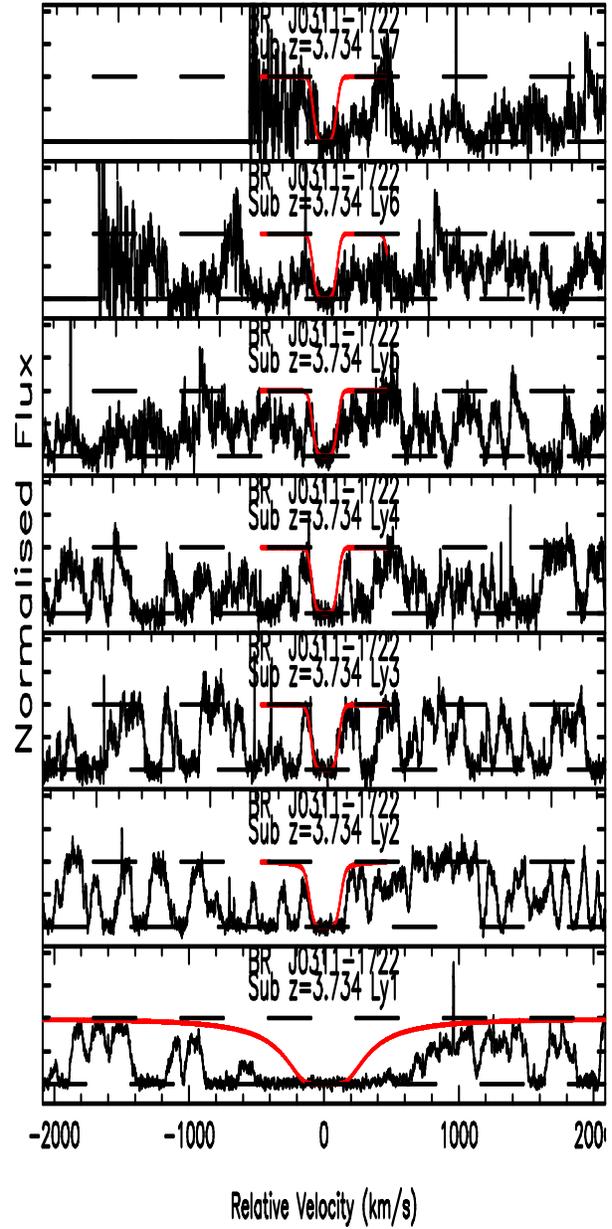}
\caption{The absorber detected towards BR J0311$-$1722 at \zabs=3.734 
(\loghi=19.48$\pm$0.10).\label{f:Q0311}} 
\end{center}
\end{figure}

\vspace{0.5cm}
\item{BR J0334$-$1612 (\zem=4.363): A DLA at \zabs=3.56 was reported by 
P\'eroux \e\ (2001) from medium resolution spectroscopy. They derive
\loghi=21.0 in excellent agreement with the measurement made here from
higher resolution spectroscopy: \loghi=21.12$\pm$0.15 with
\zabs=3.557. Nevertheless, we note the asymmetry in the shape 
of the profile, a possible signature of multi-component
absorbers. Metal lines associated with this system are detected in the
red part of the spectrum. Figure~\ref{f:Q0334} shows our best fit of
this \hi\ line. We also note in the spectrum of this quasar a real and
significant broad dip in the continuum at $\lambda \sim$ 5200
\AA. This cannot be explained by any known broad absorption lines, nor by an
ion at the emission redshift and does not correspond to the beginning
of the Lyman$\beta$ forest either. Similar features (although not as
dramatic) have been reported by Levshakov \e\ (2004). In that case,
the dips are clearly associated with \o6\ at the emission redshift and
are interpreted as ejected material or intracluster cooling
flow.}

\begin{figure}
\begin{center}
\includegraphics[height=4cm, width=8cm, angle=0]{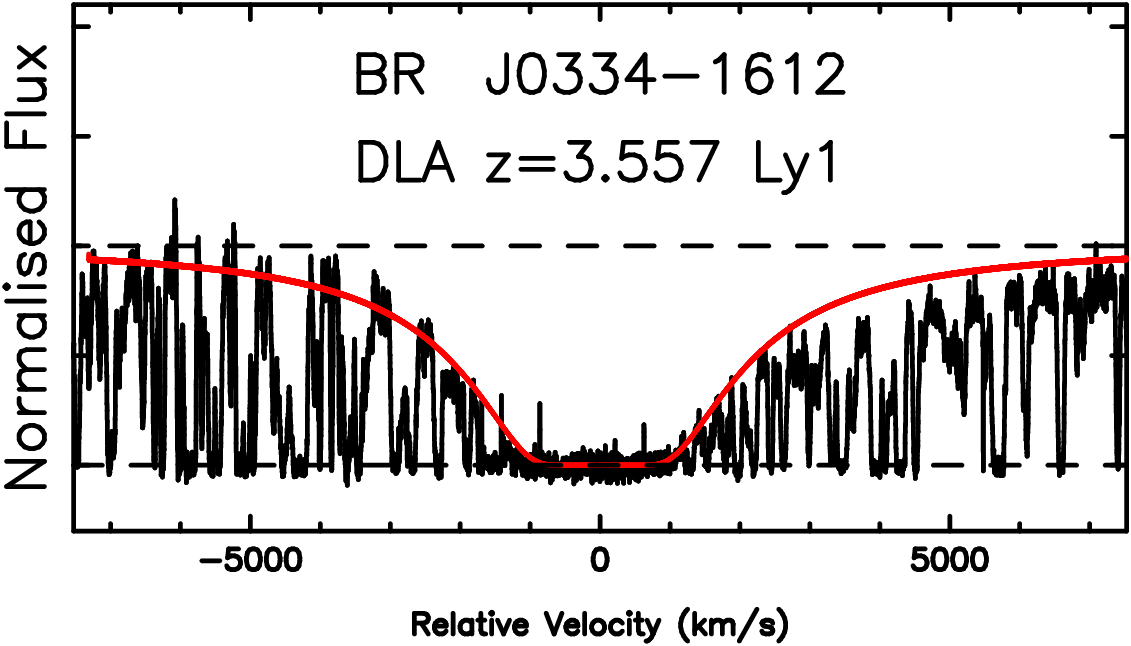}
\caption{The DLA absorber detected towards BR J0334$-$1612 with 
\zabs=3.557 (\loghi=21.12$\pm$0.15). We note the asymmetry in the shape 
of the profile, a possible signature of multi-component
absorbers. \label{f:Q0334}}
\end{center}
\end{figure}

\vspace{0.5cm}
\item{BR J0419$-$5716 (\zem=4.461): This quasar was observed by P\'eroux 
\e\ (2001) who did not detect any sub-DLA in its medium resolution spectrum. 
Using higher-resolution data, we detect a sub-DLA with \zabs=3.063
with \loghi=19.17$\pm$0.10 as well as one metal
transition. Figure~\ref{f:Q0419} shows our best fit of this \hi\
line.}

\begin{figure}
\begin{center}
\includegraphics[height=4cm, width=8cm, angle=0]{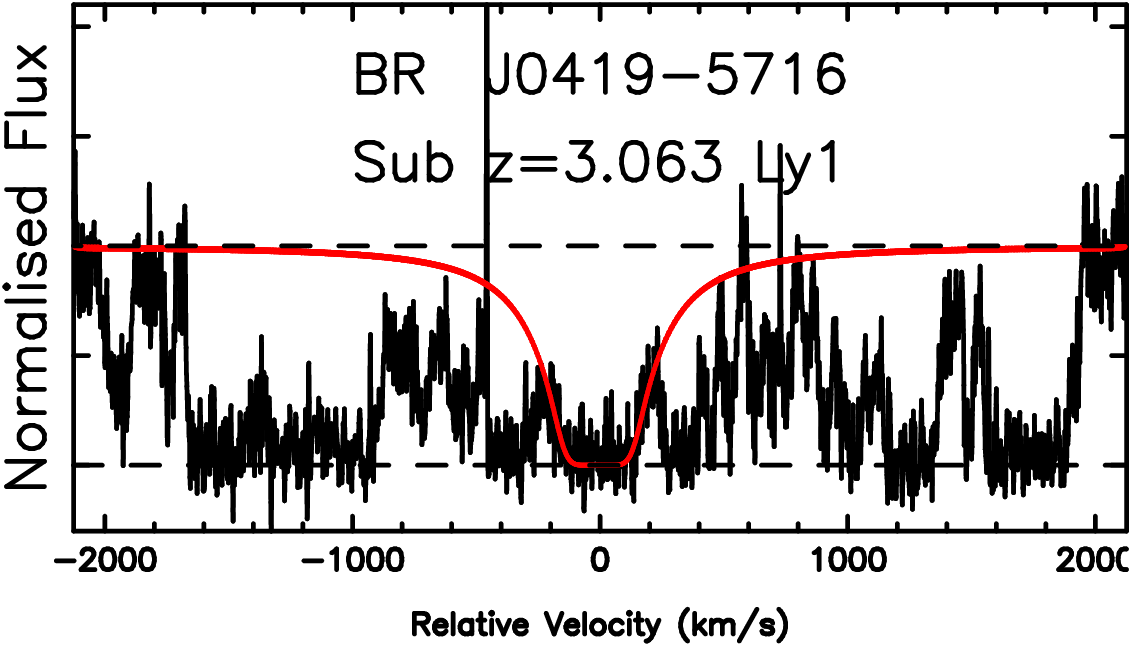}
\caption{The sub-DLA absorber detected towards BR J0419$-$5716 at 
\zabs=3.063 (\loghi=19.17$\pm$0.10).\label{f:Q0419}} 
\end{center}
\end{figure}

\vspace{0.5cm}
\item{BR J2017$-$4019 (\zem=4.131): This quasar is a Broad Absorption Line 
(BAL) quasar (P\'eroux \e\ 2001). We carefully search for DLA/sub-DLAs
in the regions free from BAL features but do not detect any. We note
however \o6\ $\lambda\lambda$1032 and 1037 features at \zabs=3.9966,
3.9692 and maybe also at \zabs=3.7300.}

\vspace{0.5cm}
\item{PSS J2154$+$0335 (\zem=4.363): This quasar was observed by P\'eroux 
\e\ (2001) who report a DLA at \zabs=3.61. This system falls in the setting 
gap of our UVES spectra. On the other hand, we discover a new sub-DLA at
\zabs=3.177 with \loghi=19.23$\pm$0.15. No metals were found associated 
with system over the limited wavelength range of our
spectrum. Figure~\ref{f:Q2154} shows our best fit of this
\hi\ line.}

\begin{figure}
\begin{center}
\includegraphics[height=4cm, width=8cm, angle=0]{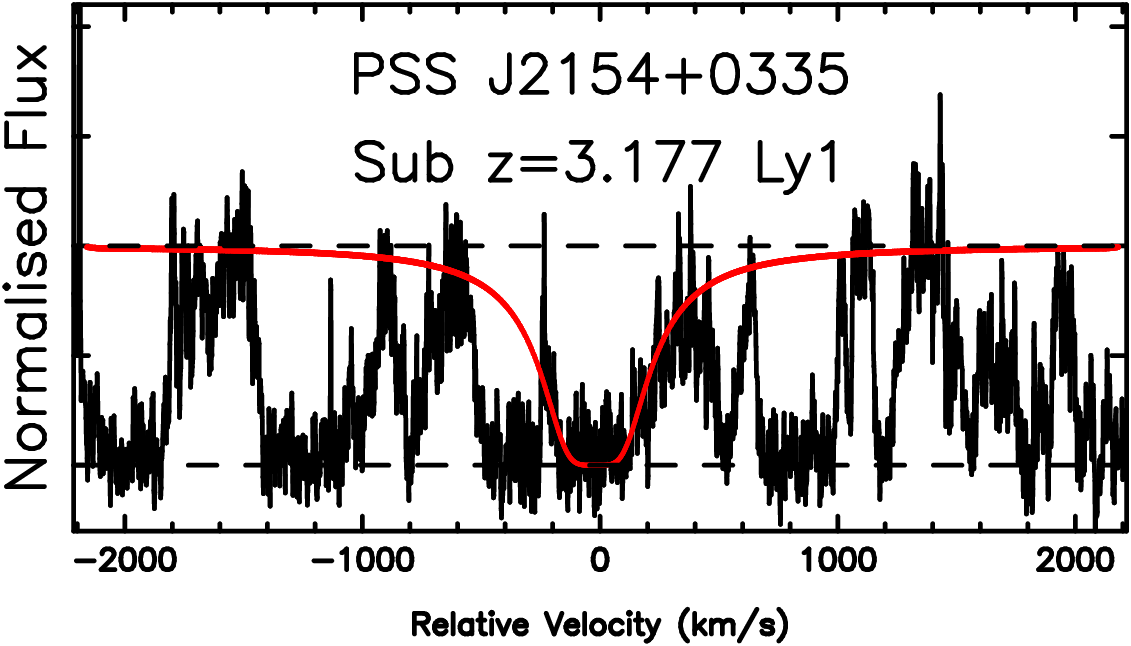}
\caption{The sub-DLA absorber detected towards PSS J2154$+$0335 at 
\zabs=3.177 (\loghi=19.23$\pm$0.15).\label{f:Q2154}} 
\end{center}
\end{figure}

\vspace{0.5cm}
\item{BR J2215$-$1611 (\zem=3.990): This quasar was observed by 
Storrie-Lombardi \e\ (1996b) who did not find any DLA in the
spectrum. Here we report on two new sub-DLAs with
\loghi=19.01$\pm$0.15 and 20.05$\pm$0.15 at \zabs=3.656 and 3.662 
respectively. The two are separated by just $\sim$320
\kms. The Lyman series down to Ly3 is detected where the division into 
two distinct systems is unambiguous. Several metal lines are detected
at these redshifts too. Figure~\ref{f:Q2215} shows our best fit of
these \hi\ lines. We also note OVI $\lambda\lambda$1032 and 1037
features at \zabs=3.9785 along this quasar line of sight.}

\begin{figure}
\begin{center}
\includegraphics[height=10cm, width=8cm, angle=0]{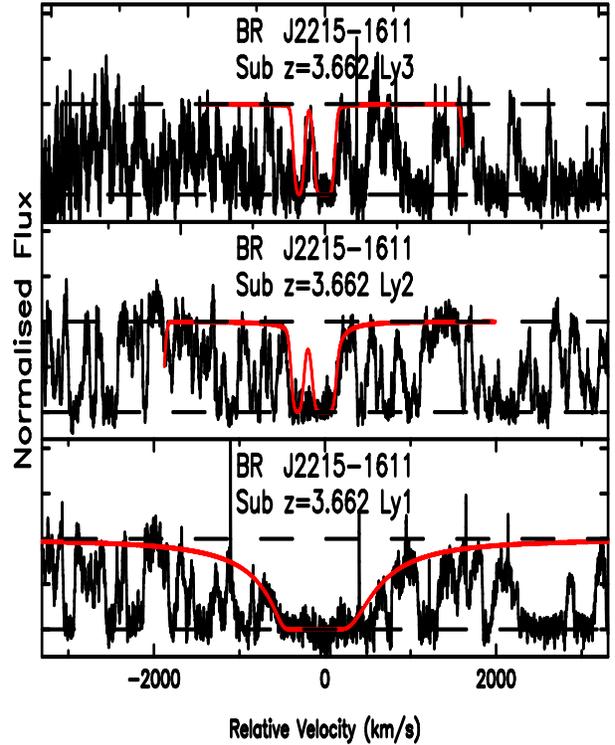}
\caption{The sub-DLA absorbers detected towards BR J2215$-$1611 at 
\zabs=3.656 (\loghi=19.01$\pm$0.15) and \zabs=3.662 (\loghi=20.05$\pm$0.15) 
respectively.\label{f:Q2215}} 
\end{center}
\end{figure}

\vspace{0.5cm}
\item{BR J2216$-$6714 (\zem=4.469): This quasar was observed by P\'eroux 
\e\ (2001) who do not report any absorber but mention a possible sub-DLA 
candidate at \zabs=3.37. We now confirm that system and derive
\loghi=19.80$\pm$0.10 at \zabs=3.368. Several metal lines are also
observed at this redshift. Figure~\ref{f:Q2216} shows our best fit of
this \hi\ line.}

\begin{figure}
\begin{center}
\includegraphics[height=4cm, width=8cm, angle=0]{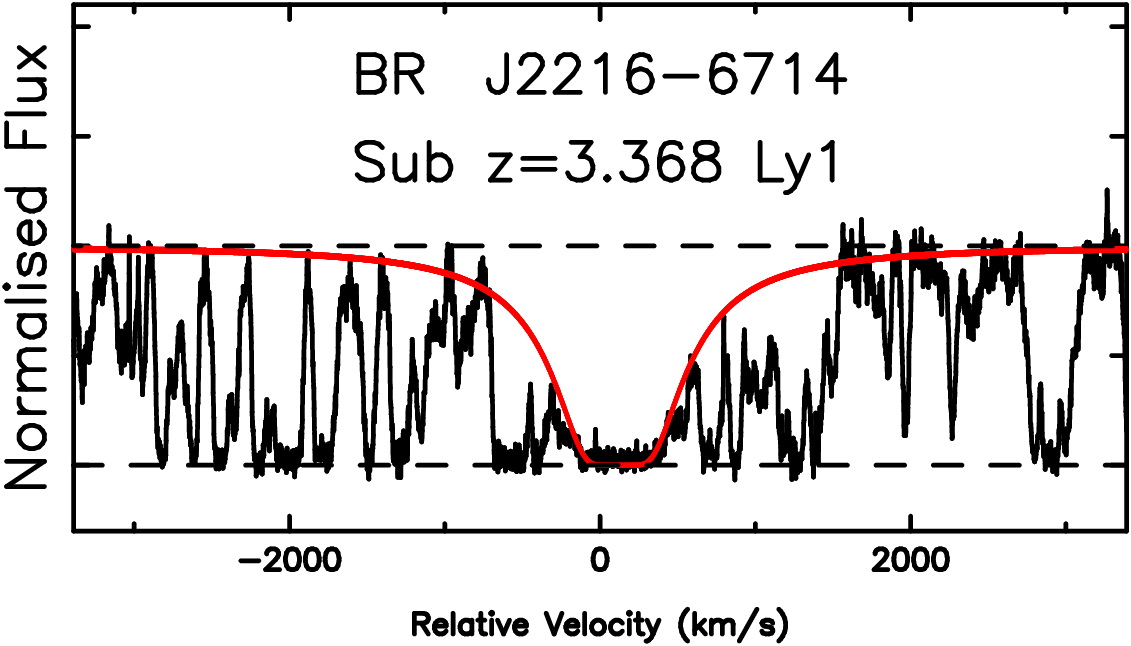}
\caption{The sub-DLA absorber detected towards BR J2216$-$6714 at 
\zabs=3.368 (\loghi=19.80$\pm$0.10).\label{f:Q2216}} 
\end{center}
\end{figure}

\vspace{0.5cm}
\item{BR J2239$-$0552 (\zem=4.558): This quasar was observed by 
Storrie-Lombardi \e\ (1996b) who report a DLA at \zabs=4.08. We now
confirm that system and derive \loghi=20.55$\pm$0.10 at
\zabs=4.079 in good agreement with estimates from medium resolution 
spectroscopy. The absorber is detected down to the Ly3 and has
associated metal lines. Figure~\ref{f:Q2239} shows our best fit of
these \hi\ lines.}

\begin{figure}
\begin{center}
\includegraphics[height=10cm, width=8cm, angle=0]{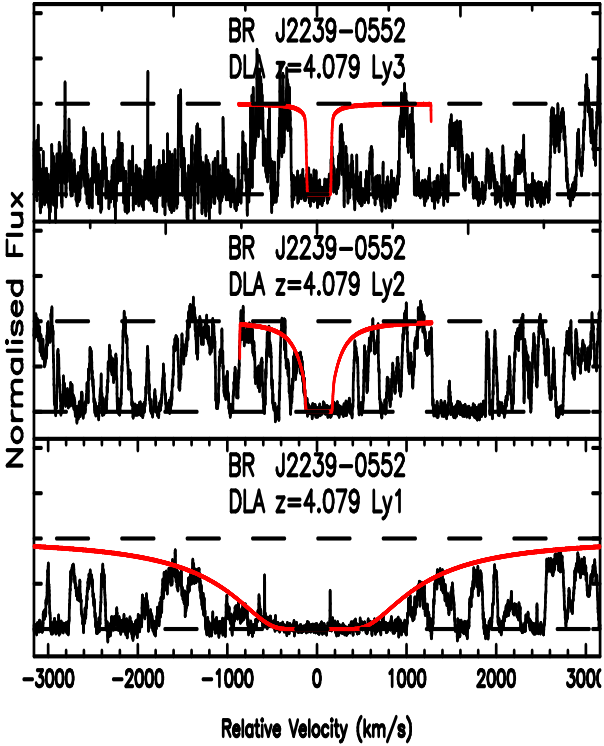}
\caption{The DLA absorber detected towards BR J2239$-$0552 at 
\zabs=4.079 (\loghi=20.55$\pm$0.10).\label{f:Q2239}} 
\end{center}
\end{figure}

\vspace{0.5cm}
\item{BR J2349$-$3712 (\zem=4.208): This quasar was observed by 
P\'eroux \e\ (2001) at medium resolution. They do not find any DLA
along its line of sight. Here, we report on the discovery of three
sub-DLAs, two of which are blended. The first system has
\loghi=19.12$\pm$0.10 at \zabs=3.581 while the two others have
\loghi=19.79$\pm$0.15 at \zabs=3.690 and \loghi=19.78$\pm$0.10 at
\zabs=3.696 (i.e. $\Delta v \sim$380\kms). The subdivision of these is 
unambiguous in Ly2. For all the systems the Lyman series is available
down to Ly3. No associated metal lines are detected over the limited
wavelength coverage of our spectrum. Figure~\ref{f:Q2349} shows our
best fit of these \hi\ lines.}

\begin{figure}
\begin{center}
\includegraphics[height=8cm, width=8cm, angle=0]{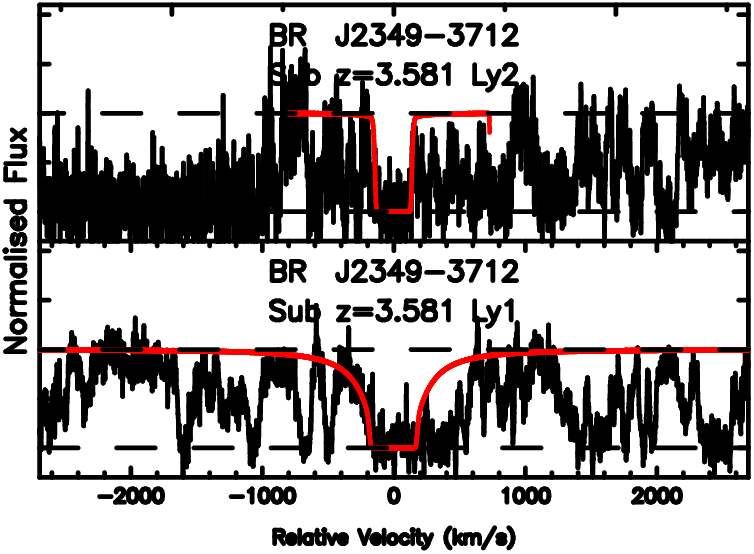}
\vspace{0.5cm}
\includegraphics[height=8cm, width=8cm, angle=0]{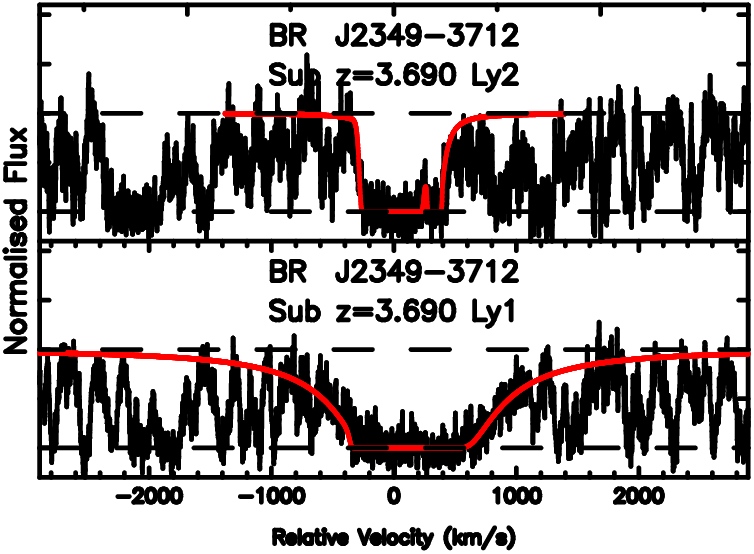}
\caption{The sub-DLA absorbers detected towards BR J2349$-$3712 at 
\zabs=3.581 (\loghi=19.12$\pm$0.10) while the two others are at 
\zabs=3.690 (\loghi=19.79$\pm$0.15) and \zabs=3.696 (\loghi=19.78$\pm$0.10).\label{f:Q2349}}
\end{center}
\end{figure}

\end{enumerate}

\subsection{Medium/High-Resolution \nhi\ Comparison}

\begin{table}
\begin{center}
\caption{This table compiles the DLA column density and redshift
estimates from 5\AA\ (FWHM), signal-to-noise ratio per pixel $\sim$ 20
(P\'eroux et~al. 2001; except BR J2239$-$0552 which is from
Storrie-Lombardi et~al. 1996b) and 2\AA, signal-to-noise ratio per
pixel $\sim$ 25 quasar spectra from the present study. \label{t:res}}
\begin{tabular}{lccccc} \hline 
Quasar & \multicolumn{2}{c}{medium resolution} &\multicolumn{2}{c}{high resolution} &$\Delta$ \nhi\\
& \zabs & \nhi & \zabs & \nhi &(med-high)\\
\hline
BR J0006$-$6208    	&3.20  &20.9  &3.202	&20.80  &$+$0.10\\
PSS J0133$+$0400        &3.69  &20.4  &3.692	&20.68	&$+$0.28\\
...                     &3.77  &20.5  &3.773	&20.42	&$-$0.08\\
PSS J0209$+$0517        &3.66  &20.3  &3.666	&20.47	&$+$0.17\\
...                     &3.86  &20.6  &3.863	&20.43  &$-$0.17\\
BR J0334$-$1612		&3.56  &21.0  &3.557	&21.12	&$+$0.12\\
BR J2239$-$0552         &4.08  &20.4  &4.079	&20.55  &$+$0.15\\
\hline
mean 		&3.69	&20.58	   &3.689      &20.64   &$+$0.08\\
min value	&3.20   &20.30	   &3.202      &20.42   &$-$0.17\\
max value	&4.08	&21.00	   &4.079      &21.12   &$+$0.28\\		
\hline
\end{tabular}
\end{center}
\end{table}
			
Seven of the quasar absorbers described above are classical DLAs with
\loghi$>$20.3. In Table~\ref{t:res}, the \nhi\ column density estimates 
from medium resolution spectroscopy of these DLAs is compared with the
new high-resolution measurements presented here. The last two systems
(toward BR J0334$-$1612 and BR J2239$-$0552) were also observed by
Storrie-Lombardi \& Wolfe (2000) who found \nhi\ in very good
agreements with ours. The comparison shows that previous \nhi\
estimates from $5$\AA\ resolution quasar spectra are reliable. In
fact, it is also known that estimates of systems with high column
density from echelle data might be affected by the difficulty to trace
the correct continuum over different echelle order. In the sub-DLA
regime however, high-resolution spectra are definitely required in
order to accurately measure
\nhi.

\section{Analysis}

\subsection{Survey's Properties}

\begin{table}
\begin{center}
\caption{Redshift path surveyed. z$_{\rm min}$ corresponds to the point 
below which no flux is observed and z$_{\rm max}$ is 3000 km s$^{\rm
-1}$ bluewards of the Ly-$\alpha$ emission line. Gap in
non-overlapping settings and known DLAs are taken into account.\label{t:gz}}
\begin{tabular}{llcc}
\hline
Quasar & z$_{\rm em}$ &z$_{\rm min}$ &z$_{\rm max}$ \\
\hline
BR J0006$-$6208    	&4.455 	&3.083  &3.188\\
... 			&... 	&3.216	&3.732\\
... 			&... 	&3.798 	&4.450\\
PSS J0118$+$0320        &4.230 	&2.767  &4.225\\
PSS J0121$+$0347        &4.127 	&2.739  &3.410\\
...                     &... 	&3.474	&4.122\\
SDSS J0124$+$0044       &3.840 	&2.414  &3.835\\
PSS J0133$+$0400        &4.154 	&2.414  &3.247\\
... 			&... 	&3.305	&3.676\\
... 			&... 	&3.707	&3.765\\
... 			&... 	&3.786	&4.149\\
BRI J0137$-$4224        &3.970 	&2.344  &3.965\\
PSS J0209$+$0517        &4.174 	&2.767  &3.658\\
... 			&... 	&3.675	&3.855\\
... 			&... 	&3.872	&4.169\\
BRI J0244$-$0134        &4.053 	&2.414  &3.247\\
... 			&... 	&3.305	&4.048\\
BR J0311$-$1722         &4.039 	&2.560  &3.410\\
... 			&... 	&3.474	&4.034\\
BR J0334$-$1612         &4.363 	&2.748  &3.541\\
... 			&... 	&3.573	&4.358\\
BR J0419$-$5716         &4.461 	&2.748  &3.572\\
... 			&... 	&3.639	&4.456\\
BR J2017$-$4019         &4.131 	&2.607  &2.994\\
... 			&... 	&3.022	&3.407\\
... 			&... 	&3.473	&4.126\\
PSS J2154$+$0335        &4.363 	&2.748  &3.572\\
... 			&... 	&3.639	&4.358\\
BR J2215$-$1611         &3.990 	&2.315  &3.408\\
... 			&... 	&3.473	&3.985\\
BR J2216$-$6714         &4.469 	&2.748  &4.464\\
BR J2239$-$0552         &4.558 	&2.894  &3.735\\
... 			&... 	&3.801	&4.070\\
... 			&... 	&4.096	&4.552\\
BR J2349$-$3712         &4.208 	&2.601  &3.407\\
... 			&... 	&3.473	&4.203\\
\hline
\end{tabular}
\end{center}			
\end{table}

{\it Survey sensitivity:} Table~\ref{t:gz} lists the minimum ($z_{\rm
min}$) and maximum ($z_{\rm max}$) redshifts along which a sub-DLA
could be detected along each quasar line-of-sight. $z_{\rm min}$
corresponds to the point below which the signal-to-noise ratio was too low
to find absorption features at the sub-DLA threshold of $W_{\rm
rest}=2.5$ \AA, and $z_{\rm max}$ is 3000 km s$^{-1}$ blueward of the
Ly$\alpha$ emission of the quasar. We took care to exclude the DLA
regions and the gaps in the spectrum due to non-overlapping UVES
settings when computing the redshift path surveyed.

\begin{figure}
\begin{center}
\includegraphics[height=7cm, width=7cm, angle=0]{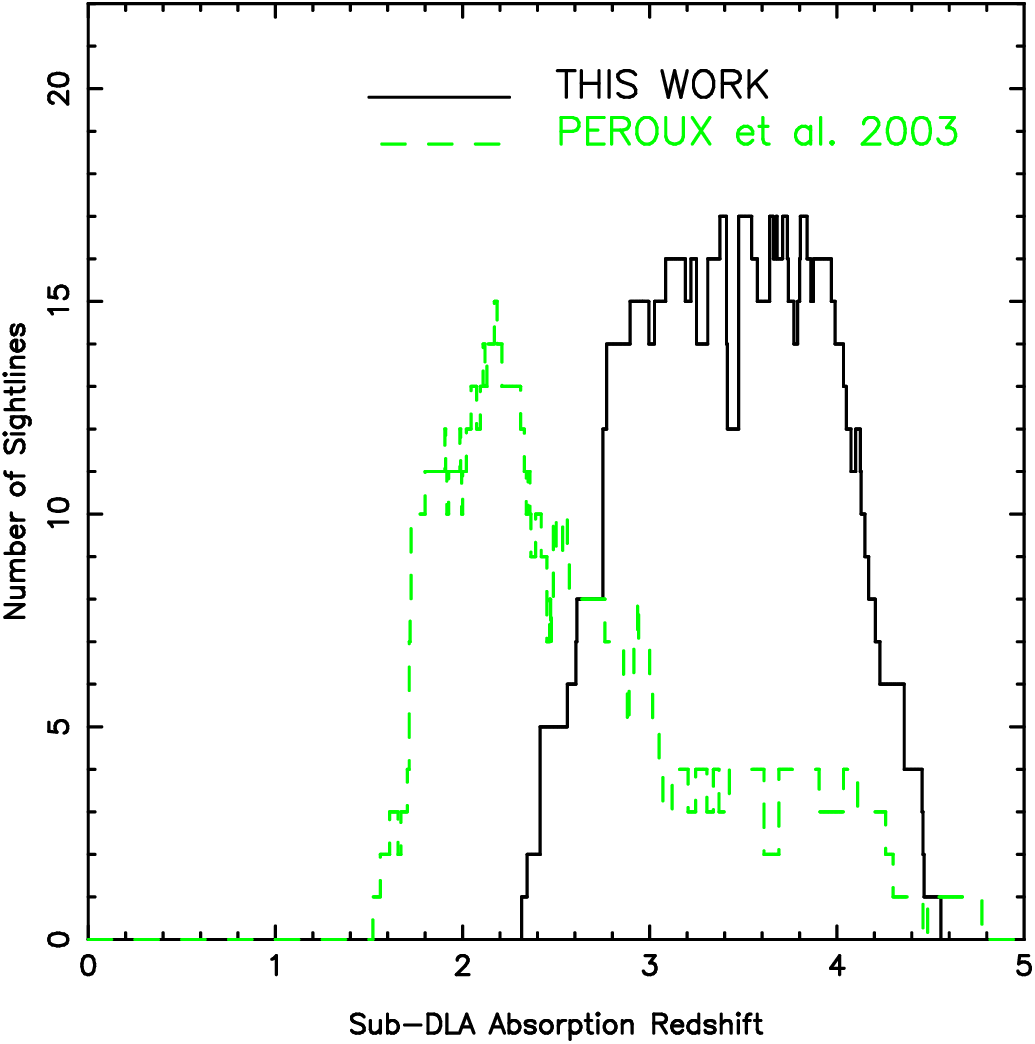}
\caption{Survey sensitivity function. The g(z)
function shows the cumulative number of lines of sight along which a
sub-DLA system could be detected. The light grey histogram represents
our archival search (\PII) while the present work is represented in
black. It shows that our new observations probe a higher redshift
interval than the archival work. The dip at $z \sim 3.45$ corresponds
to the 5 quasars for which the gap in the 540 setting was not
covered.\label{f:gz}}
\end{center}
\end{figure}

Figure~\ref{f:gz} shows the cumulative number of lines of sight along
which a sub-DLA {\it could} have been detected at the 5$\sigma$
confidence level. This survey sensitivity, $g(z)$, is defined by:

\begin{equation}
g(z) = \sum H (z^{max}_{i} - z) H (z - z^{min}_{i})
\end{equation}

where H is the Heaviside step function. In Figure~\ref{f:gz}, it is
compared with those of previous sub-DLA survey (\PII). It shows that
our new observations probe a higher redshift interval than the first
archive-based work and that the combination of the two provide a
rather homogeneous survey coverage in the range z=1.5 to 4.5.

\begin{figure}
\begin{center}
\includegraphics[height=7cm, width=5cm, angle=-90]{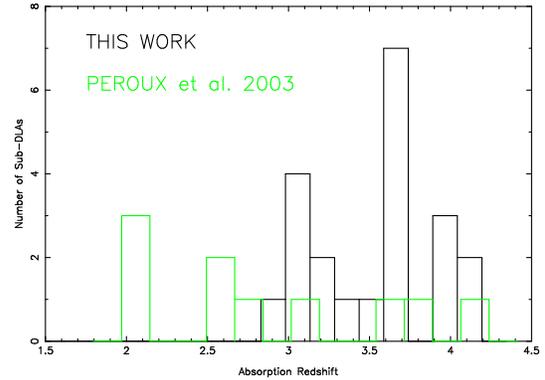}
\caption{Histogram of the redshift distribution of the sub-DLAs discovered 
as part of this survey. \PII\ refers to the statistical sample of the
archival search undertook at lower redshift (grey histogram). The
black histogram represents the new high-redshift sample from the
present study.\label{f:histo}}
\end{center}
\end{figure}

\vspace{.5cm} 
\noindent
{\it Redshift distribution:} The sub-DLA sample described in the
section above leads to 21 sub-DLAs along 17 quasar lines-of-sight
whilst we had 10 sub-DLAs making up our ``statistical sample'' toward
22 lower redshift quasars in our previous archival study (\PI). We
also confirm the presence of 7 DLAs along the same lines of sight. The
histograms showing the redshift distribution of these homogeneous
sub-DLA samples are shown in Figure~\ref{f:histo}.

\subsection{Sub-DLAs Statistical Properties}

\begin{figure}
\begin{center}
\includegraphics[height=7cm, width=7cm, angle=-90]{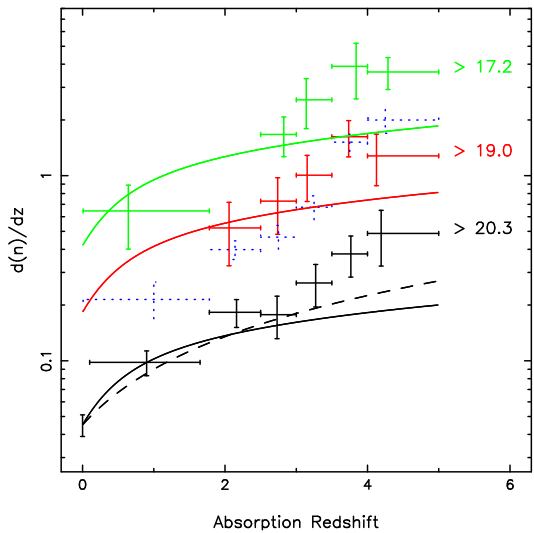}
\caption{Number density of various classes of quasar absorbers as a 
function of redshift taken from P\'eroux et al. (2003a) and Rao,
Turnshek \& Nestor (2005) except for sub-DLAs (this work). The
horizontal error bars are the bin sizes and the vertical error bars
are the 1-$\sigma$ uncertainties. The dotted bins are the predictions
of the number density of sub-DLAs from P\'eroux \e\ (2003a), while the
light grey bins at \loghi $>$19.0 correspond to the observed number
density presented in this paper. DLA's $n(z=0)$ data point is from the
21-cm emission line observations of Zwaan et al. (2005b). The curves
represent a non-evolving population for a non-zero $\Lambda-$Universe
except for the dashed line which is for a $\Lambda=0$, $q_0=0$
Universe.
\label{f:nz}}
\end{center}
\end{figure}

\begin{table*}
\begin{center}
\caption{This table summarises the observed number density of quasar absorbers 
for different column density ranges together with the predicted number
density of sub-DLAs (from P\'eroux et al. 2003a and Rao, Turnshek \&
Nestor 2005; except for the sub-DLA number density which is
from the present work). The empty entries are regions of
redshift/column density parameter space as yet unobserved.\label{t:nz}
}
\begin{tabular}{cccccccccccccc} 
\hline \hline
\loghi\ &\multicolumn{4}{c}{$>17.2$}&\multicolumn{4}{c}{$>19.0_{\rm obs}^*$}&\multicolumn{1}{c}{$>19.0_{\rm pred}$}&\multicolumn{4}{c}{$>20.3$}\\
z range &\# &$<$z$>$ &dz &n(z) &\# &$<$z$>$ &dz &n(z)&n(z)&\# &$<$z$>$
&dz &n(z)\\
\hline
0.01$-$1.78 &7  &0.64 &10.9  &0.64$\pm$0.24 &.. &...  &...   &...     	    &0.21 &n/a &0.90 &n/a  &0.10$\pm$0.02\\
1.78$-$2.50 &.. &...  &...   &...     	    &3  &2.06 &8.8   &0.52$\pm$0.20 &0.40 &34 &2.16 &186.1 &0.18$\pm$0.03\\
2.50$-$3.00 &17 &2.82 &10.2  &1.67$\pm$0.41 &5  &2.74 &9.1   &0.73$\pm$0.25 &0.46 &15 &2.73 &84.5  &0.18$\pm$0.05\\
3.00$-$3.50 &11 &3.14 &4.3   &2.57$\pm$0.77 &7  &3.15 &9.4   &1.01$\pm$0.28 &0.67 &15 &3.27 &57.0  &0.26$\pm$0.07\\
3.50$-$4.00 &9  &3.84 &2.3   &3.89$\pm$1.30 &12 &3.74 &9.6   &1.62$\pm$0.36 &1.52 &16 &3.77 &42.4  &0.38$\pm$0.09\\
4.00$-$5.00 &26 &4.29 &7.2   &3.63$\pm$0.71 &4  &4.13 &5.1   &1.28$\pm$0.40 &2.00 &9  &4.19 &18.5  &0.49$\pm$0.16\\
\hline
total       &71 &...  &34.9  &...           &31 &...  &42.0  &...           &...  &100&...  &514.8 &...\\ 
\hline \hline
\end{tabular}
\end{center}
\begin{minipage}{140mm}
{\bf \#}: number of absorption systems.\\
{\bf $19.0_{\rm obs}^*$}: \#, $<$z$>$ and dz refer to systems with 
19.0$<$\loghi$<$20.3, whilst n(z) is for all systems with \loghi$>$19.0. \\
n/a: ``not applicable'' refers to low -redshift Mg II-selected systems and so the number of 
absorption systems and corresponding dz are not directly comparable to higher-redshift statistics.
\end{minipage}
\end{table*}

{\it Redshift Number Density:} The number of quasar absorbers per unit
redshift, $n(z)$, is a direct observable. This quantity, however, can
be used to constrain the evolution or lack of it only when deconvolved
from the effect of cosmology.

The data acquired here used in combination with recent results from
the literature allow us to determine this quantity for various classes
of quasar absorbers. This is shown in Figure~\ref{f:nz} where the
number density for DLAs (P\'eroux et al. 2003a; Rao, Turnshek \&
Nestor 2005), sub-DLAs (both predictions from P\'eroux et
al. 2003a computation and new direct measurements from the present
work) and LLS are presented (also from P\'eroux et al. 2003a). It can
already be seen that the predictions overestimated the number of
sub-DLAs at $z>4$ while they underestimated the number of such systems
at $z<3.5$. The observations show that the number density of sub-DLAs
is flatter than expected. All these observations are tabulated in
Table~\ref{t:nz}. Assuming no evolution in the number density, $\Phi$
and gas cross-section $\sigma$, the lack of evolution in the redshift
number density in a non-zero $\Lambda-$Universe can be expressed as
(see for example P\'eroux et al. 2004a):

\begin{equation}
n(z)=n_0 (1+z)^2 \times \left[\frac{H(z)}{H_0}\right]^{-1}
\end{equation}

\noindent
where

\begin{equation}
\frac{H(z)}{H_0} = \left[\Omega_Mz(1+z)^2-\Omega_{\Lambda}[z(z+2)]+(1+z)^2\right]^{1/2}
\label{e:hz}
\end{equation}

These no evolution curves are shown for each class of quasar absorbers
in Figure~\ref{f:nz} where $n_0$ is taken to be $n(z=0)=0.045\pm0.006$
derived by Zwaan et al. (2005b) from high-resolution 21-cm emission
line observations of the $z=0$ analogs of DLAs. The other curves are
scaled according to a factor representative of the difference in
redshift number density of sub-DLAs and LLS in the redshift range
$2.5<z<3.0$. Departure from no evolution at $z\geq3$ is clear for all
classes of quasar absorbers. For comparison with previous work, the
dashed line shows the non-evolving population for a $\Lambda=0$,
$q_0=0$ Universe.

\begin{figure}
\begin{center}
\includegraphics[height=7cm, width=7cm, angle=-90]{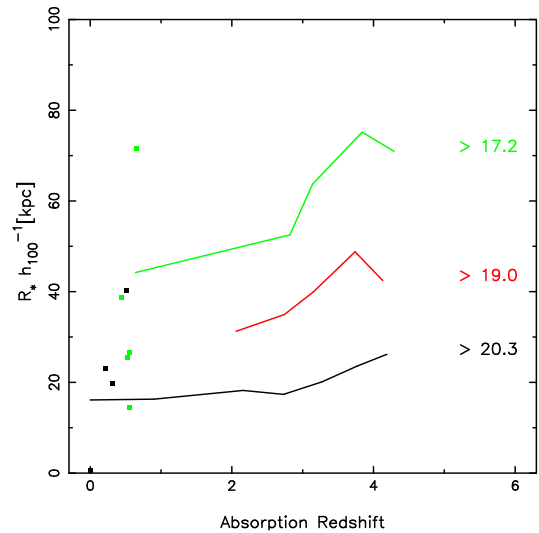}
\caption{The redshift
evolution of the characteristic radius, $R_*$, of DLAs, sub-DLAs and
LLS. For comparison, the impact parameters of spectroscopically
confirmed $z<1$ DLAs (see Boissier, P\'eroux \& Pettini 2003 for a
compilation) and MgII-selected galaxies, a proxy for LLS, (Steidel et
al. 2002) are also plotted.
\label{f:abs_size}}
\end{center}
\end{figure}

\vspace{.5cm} 
\noindent
{\it Size of Quasar Absorbers:} With the assumption that all classes
of absorbers (i.e. DLAs, sub-DLAs and LLS) and all members of a given
class are arising from the same underlying parent population, one can
estimate the cross-section radius of spherical absorbers, $\sigma$,
from the observed redshift number density $n(z)$. Following Tytler
(1981) and using a non-zero $\Lambda$-cosmology, we find:

\begin{equation}
n(z)=\frac{c}{H_0}(1+z)^2 \left[\frac{H(z)}{H_0}\right]^{-1} \int_{0}^{\infty} \epsilon \Phi(L) \kappa \pi R^2(L) d(L)
\label{e:nz}
\end{equation}

\noindent
where $\Omega_M$ is the matter density, $\Omega_{\Lambda}$ is the
contribution of the cosmological constant, $R(L)$ is the average \hi\
absorption cross-section radius of a spherical galaxy with luminosity
$L$, $\epsilon$ is the fraction of galaxies which have gaseous
absorbing envelopes with filling factor $\kappa$. $\Phi(L)$ is
estimated from the Schechter (1976) galaxy luminosity function:

\begin{equation}
\Phi \left(\frac{L}{L_*}\right)= \Phi_*\left(\frac{L}{L_*}\right)^{-s} exp \left(\frac{L}{L_*}\right)
\end{equation}

\noindent
If we further assume that a Holmberg (1975) relation between optical
radius and luminosity holds at all redshifts

\begin{equation}
\frac{R}{R_*}=\left(\frac{L}{L_*}\right)^t
\end{equation}

\noindent
we obtain the following relation for the radius:

\begin{equation}
R_*^{-2}=\frac{c}{H_0}\frac{(1+z)^2}{n(z)}\left[\frac{H(z)}{H_0}\right]^{-1} \epsilon \Phi_* \kappa \pi \Gamma(1+2t-s) 
\end{equation}

\noindent
where $\Gamma(x)$ is the Gamma function. We take $\epsilon$=1,
$\kappa$=1 and $t$=0.4 (derived from Peterson \e\ 1979). Two different
sets of parameters are used for the luminosity function:
$\Phi_*=0.0149\pm0.04$ h$^3$ Mpc$^{-3}$ and $s=1.05\pm0.01$ for
$z<0.75$ from the Sloan measurements of Blanton et al. (2003) and
$\Phi_*=0.0142^{+0.0015}_{-0.0013}$ h$^3$ Mpc$^{-3}$ and
$s=0.50^{+0.08}_{-0.06}$ for $z>0.75$ from infra-red selected galaxies
of Chen et al. (2003). These yield a \hi\ gas radius of:

\begin{equation}
R_*= A h_{100}^{-1} \frac{n(z)^{1/2}}{(1+z)} \left[\frac{H(z)}{H_0}\right]^{1/2}{\rm [kpc]}
\end{equation}

\noindent
where A=76 for $z<0.75$ and A=91 for $z>0.75$. The redshift evolution
of the characteristic radius of DLAs, sub-DLAs and LLS are plotted in
Figure~\ref{f:abs_size}. All classes of absorbers have their
characteristic radius, $R_*$, decreasing with decreasing redshift, and
this effect is stronger at lower column densities. For comparison, the
impact parameters of spectroscopically confirmed $z<1$ DLAs (see
Boissier, P\'eroux \& Pettini 2003 for a compilation) and
MgII-selected galaxies, a proxy for LLS, (Steidel et al. 2002) are
also plotted. Given the number of assumptions made in the calculation,
the few available observations are in relatively good agreement with
our results.

\begin{figure*}
\begin{center}
\includegraphics[height=7cm, angle=0]{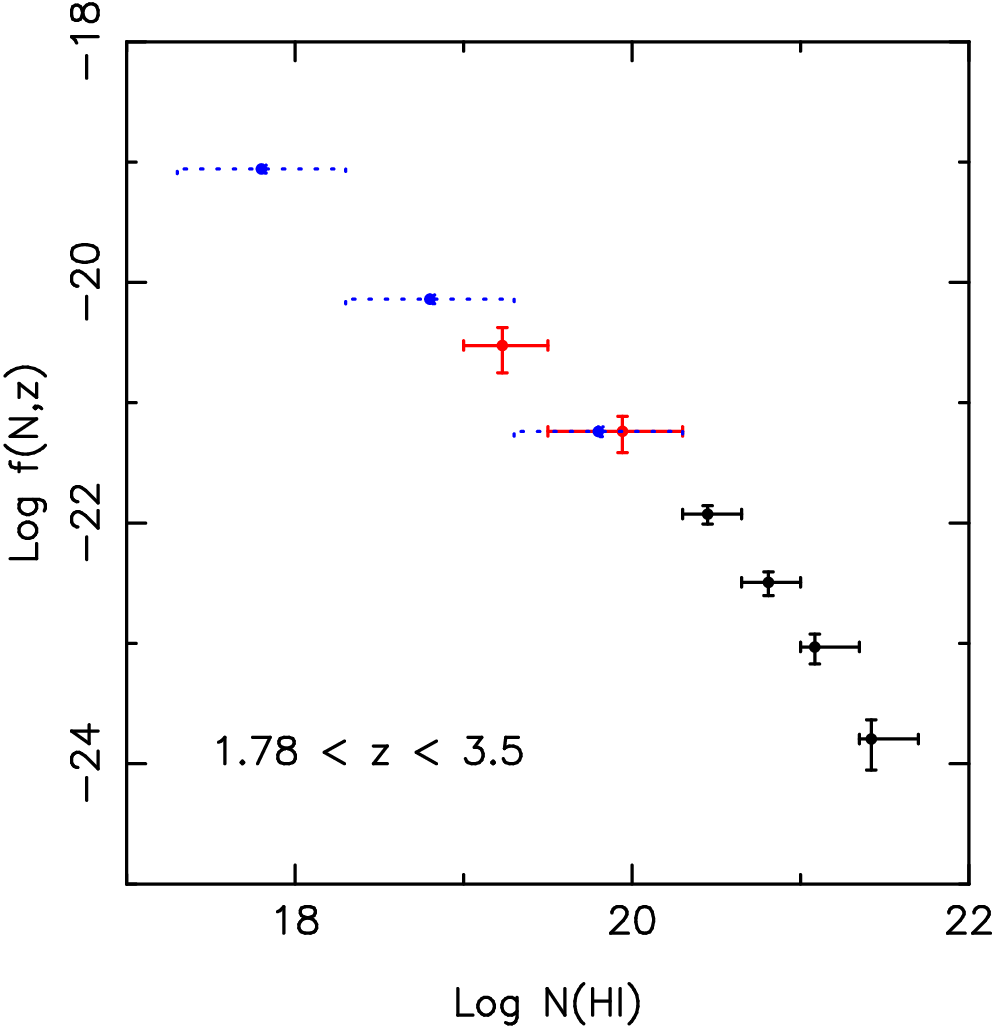}
\includegraphics[height=7cm, angle=0]{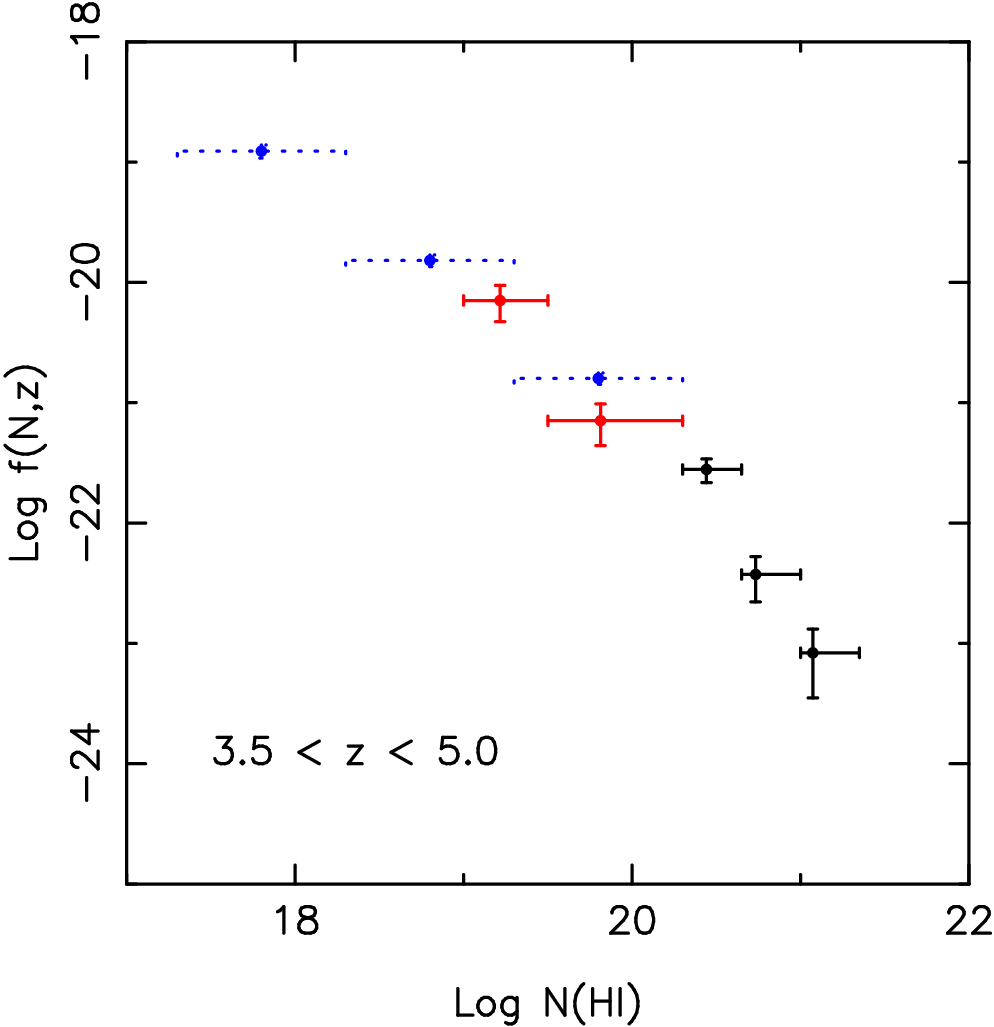}
\caption{Column density distributions for two redshift ranges down to 
the sub-DLA definition. The horizontal error bars are the bin sizes
and the vertical error bars represent the uncertainties. The dotted
bins are the predictions from fitting a Schechter function to the
expected number of LLS (P\'eroux \e\ 2003a), while the two solid bins
at 19.0$<$\loghi $<$20.3 correspond to the direct observations from
the sample of sub-DLAs presented here. \label{f:fn}}
\end{center}
\end{figure*}

\begin{table*}
\begin{center}
\caption{This table summarises the observed column density distribution 
of quasar absorbers down to the sub-DLA column density. The empty
entry in the highest column density bin at the highest redshift range
reflects the lack of systems with \loghi$>$21.35 at z$>$3.5.\label{t:fn}}
\begin{tabular}{ccccccccccc} 
\hline \hline
z range &\multicolumn{5}{c}{1.78$-$3.5}&\multicolumn{5}{c}{3.5$-$5.0}\\
\loghi&&$\log$&$\log$&$\log$&$\log$&&$\log$&$\log$&$\log$&$\log$\\
 range &\# & $<$\nhi$>$ & f(N,z)& f$_{min}$(N,z)& f$_{max}$(N,z)&\# & $<$\nhi$>$ & f(N,z)& f$_{min}$(N,z)& f$_{max}$(N,z)\\
\hline
19.00$-$19.50  & 6  &19.23  &$-$20.5 &$-$20.4 &$-$20.8 & 9  &19.22  &$-$20.2 &$-$20.0 &$-$20.3\\
19.50$-$20.30  & 9  &19.94  &$-$21.2 &$-$21.1 &$-$21.4 & 7  &19.81  &$-$21.1 &$-$21.0 &$-$21.4\\
20.30$-$20.65  &33  &20.45  &$-$21.9 &$-$21.9 &$-$22.0 &20  &20.44  &$-$21.6 &$-$21.5 &$-$21.7\\
20.65$-$21.00  &20  &20.81  &$-$22.5 &$-$22.4 &$-$22.6 & 6  &20.73  &$-$22.4 &$-$22.3 &$-$22.7\\
21.00$-$21.35  &13  &21.08  &$-$23.0 &$-$22.9 &$-$23.2 & 3  &21.07  &$-$23.1 &$-$22.9 &$-$23.5\\
21.35$-$21.70  & 5  &21.42  &$-$23.8 &$-$23.6 &$-$24.1 & 0  &...    &...\\  
\hline \hline
\end{tabular}
\end{center}
\end{table*}

\vspace{.5cm} 
\noindent
{\it Column Density Distribution:} The differential column density
distribution describes the evolution of quasar absorbers as a function
of column density and redshift. It is defined as:

\begin{equation}
f(N, z) dN dX = \frac{n}{\Delta N \sum_{i=1}^{m} \Delta X_i} dN dX
\end{equation}

where $n$ is the number of quasar absorbers observed in a column
density bin $[N, N+\Delta N]$ obtained from the observation of $m$
quasar spectra with total absorption distance coverage $\sum_{i=1}^{m}
\Delta X_i$. The column density distribution for two redshift ranges, 
z$<$3.5 and z$>$3.5 are shown in Figure~\ref{f:fn}. These redshifts
are chosen to allow a direct comparison with the work of P\'eroux \e\
(2003a). In particular, the highest redshift is set to z$=$5 to match
DLA surveys even though the sub-DLA search is made only up to z$=$4.5
(see Figure~\ref{f:gz}). Again, the observations (solid bins) are
compared with the predictions from P\'eroux \e\ (2003a). The new data
allow us to determine $f(N,z)$ down to \loghi=19.0. The column density
distribution has often been fitted with a simple power law
(i.e. Prochaska \& Herbert-Fort 2004), but there have been recent work
showing that a Schechter-type of function is more appropriate (Pei \&
Fall 1995; Storrie-Lombardi, McMahon \& Irwin 1996b; P\'eroux et
al. 2003a). In any case, it should be noted that {\it none} of these
two functions is physically motivated but rather are chosen so as to
best describe the data. However, as more observations are available, a
clear departure from the power law is observed. This flattening of the
distribution in the sub-DLA regime is indeed expected as the quasar
absorbers become less self-shielded and part of their neutral gas is
being ionised by incident UV flux. We note once more the paucity of
very high column density DLAs at high-redshift. All the data are
summarised in Table~\ref{t:fn}.

\subsection{$\Omega_{\rm HI+HeII}$ Gas Mass}

\begin{figure}
\begin{center}
\includegraphics[height=7cm,angle=-90]{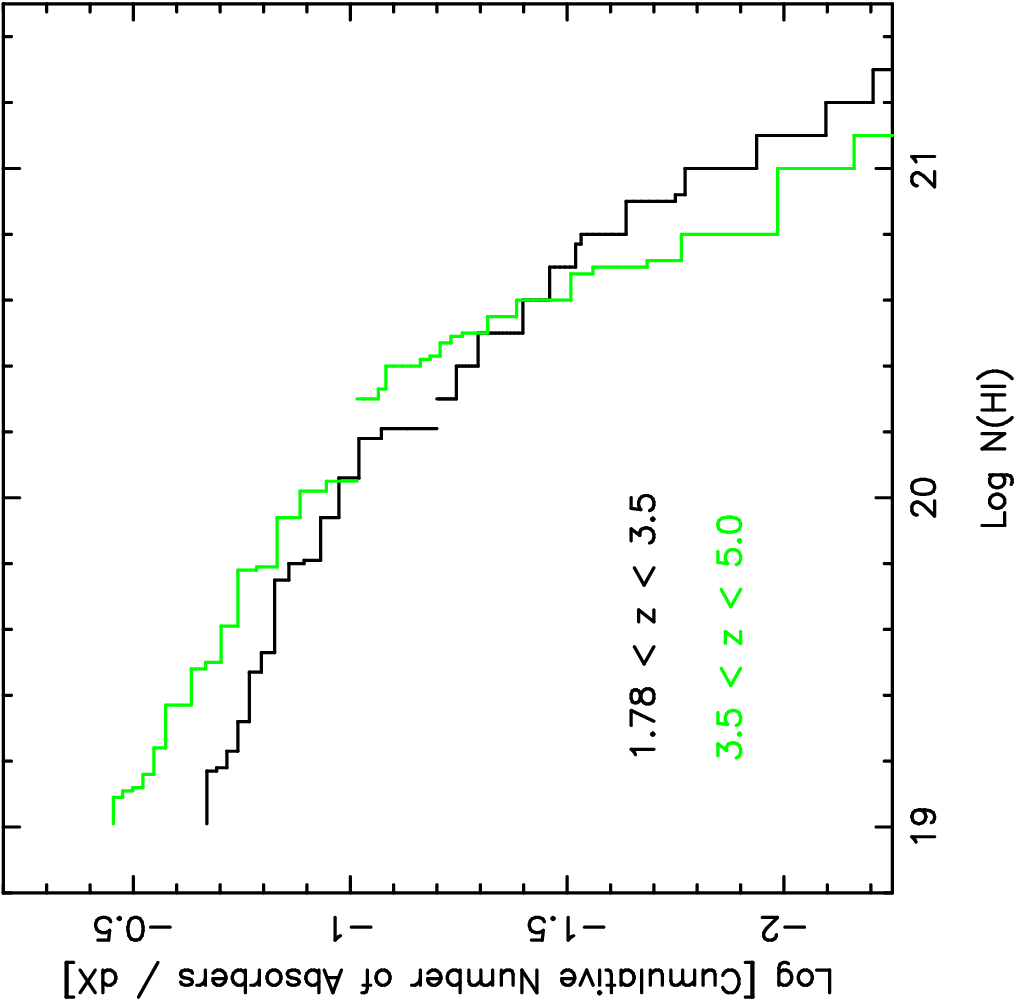}
\caption{Cumulative number of quasar absorbers as a function of column density 
for two redshift intervals. The discontinuity at \loghi=20.3
illustrate that the DLA and sub-DLA samples (and associated redshift
paths) are made of different quasar spectra and are therefore totally
{\it independent}. The observations show that the incidence of low
column density absorbers is bigger at high redshift, as predicted by
P\'eroux \e\ (2003a).\label{f:cumu} }
\end{center}
\end{figure}

{\it Cumulative Number of Absorbers:} In Figure~\ref{f:cumu}, the
number of quasar absorbers per unit comoving distance interval is
plotted as a function of \loghi. The new observations increase by one
dex the \nhi\ column density parameter space probed and therefore
provides a much better handle on the true shape of the
distribution. This, in turns, is crucial to constrain the total
\hi\ gas content which is a function of both \nhi\ column densities
and redshifts. In our surveys, the DLA and sub-DLA samples (and
associated redshift paths) are totally {\it independent}
(Figure~\ref{f:cumu}) in the sense that the quasars used to make up
the samples are different in both cases given that higher resolution
is required for the study of sub-DLAs. Therefore, the shape of both
sections of the curves are unrelated and the fact that the function in
the sub-DLA regime is a smooth continuation of the DLA part is
resulting from the observations.

\begin{table*}
\begin{center}
\caption{Redshift evolution of the gas mass density 
$\Omega_{\rm HI+HeII}$ contained in classical DLAs and in sub-DLAs and
expressed as a fraction of today's critical density. The total
amount of \hi\ gas is given {\bf in bold}.\label{t:omega_dla}}
\begin{tabular}{cccccccccccccc} 
\hline \hline
\loghi\ range &\multicolumn{6}{c}{\loghi $>$ 20.3}&\multicolumn{6}{c}{19.0$<$ \loghi $<$ 20.3}\\
 &    &    & &   $\Omega_{DLA} $ &\# &\#&& & & $\Omega_{sub-DLA}$ &\#  &\#& {\bf Total}\\
z range &$<$z$>$    &dz    &dX &   $\times 10^{-3}$ &DLA &quasar&$<$z$>$ &dz &dX &$\times 10^{-3}$ &sub-DLA &quasar& {\bf $\Omega_{\rm HI+HeII}$}\\
\hline
1.78$-$2.50  &2.17  &189.0 &586.0  &0.80$\pm$0.22 &36   &417    &2.06    &8.8 &27.6  &0.19$\pm$0.13 & 3    &35 &{\bf 0.99$\pm$0.22}\\
2.50$-$3.00  &2.73  & 87.2 &301.0  &1.02$\pm$0.31 &18   &274  	&2.74    &9.1 &31.5  &0.17$\pm$0.10 & 5    &28 &{\bf 1.18$\pm$0.31}\\
3.00$-$3.50  &3.25  & 64.8 &239.6  &1.07$\pm$0.31 &18   &178  	&3.15    &9.4 &34.9  &0.22$\pm$0.11 & 7    &35 &{\bf 1.32$\pm$0.31}\\
3.50$-$4.00  &3.77  & 49.5 &194.6  &0.80$\pm$0.23 &19   &112  	&3.74    &9.6 &37.9  &0.23$\pm$0.08 &12    &32 &{\bf 1.03$\pm$0.23}\\
4.00$-$4.99  &4.20  & 23.2 & 95.6  &0.73$\pm$0.25 &10   & 82  	&4.13    &5.1 &20.9  &0.16$\pm$0.10 & 4    &20 &{\bf 0.89$\pm$0.25}\\
\hline \hline
\end{tabular}
\end{center}
\end{table*}

\begin{figure*}
\begin{center}
\includegraphics[height=7cm,angle=0]{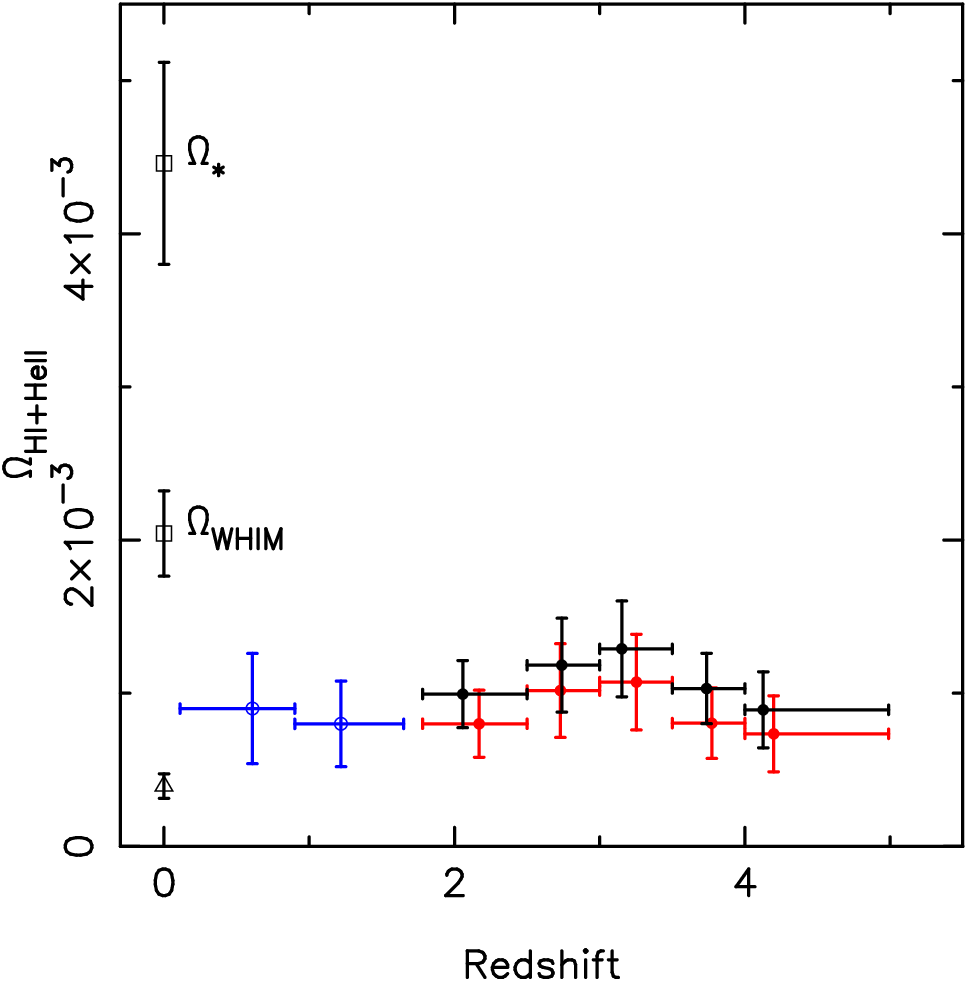}
\includegraphics[height=7cm,angle=0]{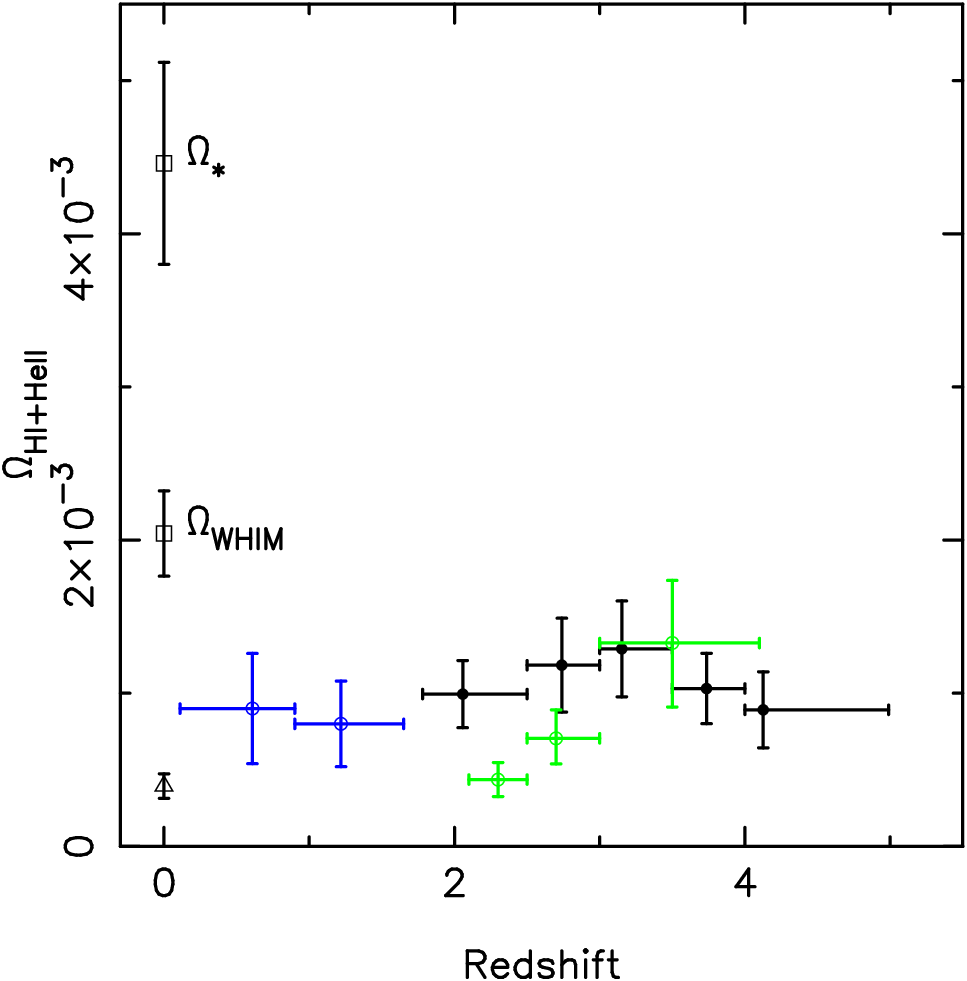}
\caption{Redshift evolution of the gas mass density $\Omega_{\rm HI+HeII}$ 
expressed as a fraction of the critical density. The stellar mass
density today is represented by $\Omega_{*}$ measured by Cole \e\
(2001) for a Salpeter Initial Mass Function and the baryons in the hot
gas is represented by $\Omega_{\rm WHIM}$ measured from the incident
of OVI absorbers by Danforth \& Shull (2005). The triangle at z=0
corresponds to the \hi\ mass measured with radio observations in local
galaxies (Zwaan \e\ 2005a). The two error bars at z$<$2 are
$\Omega_{\rm HI+HeII}$ from MgII-selected DLAs (Rao, Turnshek \&
Nestor 2005). {\bf Left panel:} the light grey bins at z$>$2 are
$\Omega_{\rm HI+HeII}$ in systems with \loghi$>$20.3, while the black
bins correspond to the total \hi\ gas, including the \hi\ contained in
sub-DLAs. {\bf Right panel:} $\Omega_{\rm HI+HeII}$ derived from the
first Data Release of the Sloan survey is shown in light grey
(Prochaska \& Herbert-Fort 2004) together with our measurements of the
total as mass (black bins).
\label{f:omega_dla}
}
\end{center}
\end{figure*}

\vspace{.5cm} 
\noindent
{\it $\Omega_{\rm HI+HeII}$:} The gas mass density, $\Omega_{\rm
HI+HeII}$, observed in high-redshift quasar absorbers is classically
expressed as a fraction of today's critical density:

\begin{equation}
\Omega_{HI}(z) = \frac{H_o \mu m_H}{c \rho_{crit}} \frac{\sum N_i(HI)}{\Delta X}
\end{equation}

where $\mu=1.3$ is the mean molecular weight and $m_H$ is the hydrogen
mass. The total gas mass, including \hi\ gas in systems below the
canonical DLA definition are plotted in Figure~\ref{f:omega_dla} and
compared with the stellar mass density today ($\Omega_{*}$) and
\hi\ mass measured with radio observations of local galaxies (triangle
at z=0). Measurements at z$<$2, are $\Omega_{\rm HI+HeII}$ from
MgII-selected DLAs (Rao, Turnshek \& Nestor 2005). The left panel of the
figure decomposes the \hi\ mass contained in both classical DLAs and
in DLAs + sub-DLAs. These results are tabulated in
Table~\ref{t:omega_dla}. The right panel of the same figure shows
$\Omega_{\rm HI+HeII}$ derived from the homogeneous survey of DLAs found in
the first Data Release (DR1) of the Sloan Sky Digital Survey
(Prochaska \& Herbert-Fort 2004).

\subsection{Clustering Properties}

As already noted in \PII, quite a few of the 21 sub-DLAs which make up
the new sample are close in redshift. The notes in the last column of
table~\ref{t:HI} emphasises in particular these systems less than
$\sim$400 \kms apart, which in practice means that they needed to be
fitted together: \zabs=3.995$/$3.999 toward PSS J0133$+$0400 ($\Delta
v=240$ \kms), \zabs=3.656$/$3.662 toward BR J2215$-$1611 ($\Delta
v=380$ \kms) and \zabs=3.690$/$3.696 toward BR J2349$-$3712 (also have
$\Delta v=380$ \kms). Another example is the two
\zabs=3.692$/$3.773 DLAs along PSS J0133$+$0400 ($\Delta v=5100$ \kms),
with in between a \zabs=3.760 quasar absorber just below the sub-DLA
definition \loghi$<$19.0. In fact, the line of sight toward PSS
J0133$+$0400 is particularly rich, containing a total of 6
DLAs/sub-DLAs, all of which but one being separated by no more than
$\Delta v=15000$ \kms. This complex group of systems appears to be
alike the multiple DLAs (MDLAs) studied in Lopez \& Ellison (2003) and
which are found to have abundance patterns distinctly different from
the one of classical DLAs. In the present data, we do not cover the
red part of the spectrum which would allow to detect metal lines to
test further the hypothesis from Lopez \& Ellison (2003). Acquiring a
spectrum covering the red part of this quasar would allow to test
further the hypothesis about the possible truncated star
formation of these MDLAs.

\section{Discussion}

\subsection{On the Ionized Fraction of Sub-DLAs}

The extension of the classical DLA definition to \loghi\ $>$ 19
proposed by P\'eroux \e\ (2003a) has triggered concerns about the
ionized fraction in these sub-DLAs. This issue is not relevant when
quasar absorbers are counted up in order to quantify the total \hi\
gas mass of the Universe, $\Omega_{\rm HI+HeII}$. Indeed, it is the
sheer number of sub-DLAs which makes them add up to the classical DLA
contribution, and this regardless of the amount of gas ionized in
these systems. On the contrary, the quasar absorbers which have a high
ionization fraction should not be included when measuring the {\it
neutral} gas mass, as they trace hotter gas. Nevertheless, the
evolution of neutral gas mass most probably is the product of several
phenomena including star formation, galactic feedback, ionization of
neutral gas as well as formation of molecular $H_2$ gas (which will
in turn leads to the formation of stars). Which of these are the
dominant processes probed by DLAs and/or sub-DLAs is at present
unclear.

When sub-DLAs are used for estimating the global metallicity, one
might rightly wonder about the fraction of the gas (and consequently
metals) in ionized form. In order to address this question, \PI\ have
carefully modelled every one of the 12 sub-DLAs making up their sample
using detailed CLOUDY model. Overall, their findings demonstrate that
indeed the ionized fraction $x=H^+/H$ account for 2/3$^{\rm rd}$ of
the gas. But they have also shown that relative or absolute abundances
estimated from $X^+/H^o$ measurements are a good approximation of
i.e. $X_{tot}/Fe_{tot}$ or $X_{tot}/H_{tot}$. Therefore, using the
metals commonly detected in sub-DLAs {\it for a measure of the
metallicity of the \hi\ gas in the Universe does not introduce a
systematic bias}, even though the ionization fraction of the systems
is not small.

\subsection{On the Nature of Sub-DLAs}

By assuming that both DLAs and sub-DLAs trace the same underlying
parent population, a natural explanation for the nature of sub-DLAs
could be that they are the outermost part of galaxies (\PII). This is
illustrated by the absorber sizes calculations presented here where
the characteristic radius of sub-DLAs is around 40 h$_{100}^{-1}$ kpc
and the one from DLAs is 20 h$_{100}^{-1}$ kpc.

The metallicity of sub-DLAs also seem to differ from the one of
classical DLAs. Smoothed particle hydrodynamics simulations (Nagamine,
Springel \& Hernquist 2004a; Nagamine, Springel \& Hernquist, 2004b)
indicate that DLAs should be 1/3$^{\rm rd}$ solar at $z=2.5$ and even
more metal-rich toward lower redshifts. Indeed there are lines of
evidence pointing toward lower column density quasar absorbers like
sub-DLAs being more metal rich at $z<2$ (P\'eroux
\e\ 2003b). This could be explained by classical DLAs being dustier
than their sub-DLAs counterparts thus providing for the selection of
against their background quasar in the first place. Recent
computations have shown that even current radio-selected quasar
samples (Ellison \e\ 2001) are not in disagreement with such a
scenario (Vladilo \& P\'eroux 2005). In fact, from figure 9a of
Vladilo \& P\'eroux (2005), it can be appreciated that in the decade
\loghi=20$-$21, from 10 to 50\% of the DLAs might be missed due to
dust obscuration. Extrapolations of these mathematical formulations to
the \loghi=19$-$20 decade suggest that only less than 10\% of the
sub-DLAs might be missed by dust obscuration.

If confirmed, this can be explained by the fact that in sub-DLAs,
the Zn column density threshold does not combine with the HI threshold
\loghi$>$20.3 that prevent their detection.

\subsection{Cosmological Evolution of the Mass Densities}
 
\begin{figure}
\begin{center}
\includegraphics[height=7cm, width=7cm, angle=-90]{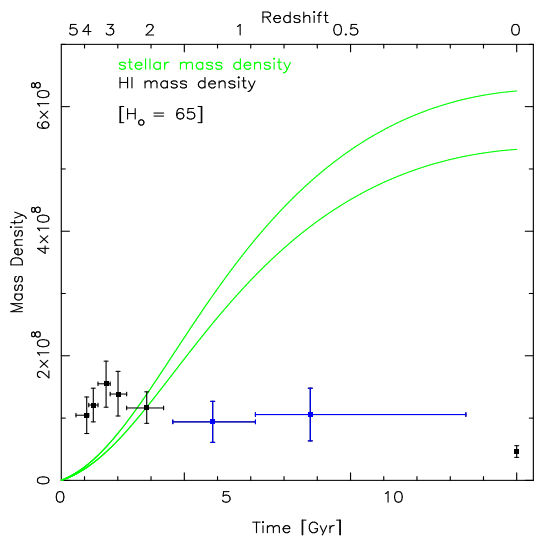}
\caption{Observable baryons in the Universe as a function of time. The 
curve represents the mass density in stars from Rudnick \e\ (2003)
integrated from the Star Formation Rate measured by Cole \e\
(2001). The uncertainties in that measurement are estimated to be
around 15\% but are also a strong function of the chosen Initial Mass
Function. The error bars represent the mass density in the quasar
absorbers deconvolved from the local critical density. A direct
comparison of these two quantities illustrates the puzzling current
situation concerning the baryon content of the
Universe.\label{f:omega_all}}
\end{center}
\end{figure}

Figure~\ref{f:omega_all} shows the cosmological evolution of some of
the observable baryons in the Universe. In recent years, new
observations have considerably changed the global picture with respect
to previous studies (Lanzetta \e\ 1991; Wolfe \e\ 1995;
Storrie-Lombardi
\e\ 1996a; Storrie-Lombardi \& Wolfe 2000). In Figure~\ref{f:omega_all}, 
the direct comparison of the mass density in stars and in \hi\
illustrate several issues:

\begin{enumerate}

\item $\Omega_{\rm HI+HeII}$ observed in quasar absorbers at high redshift 
($z>2$) is low compare with the mass density observed in stars today,
$\Omega_{*}$.

\item $\Omega_{\rm HI+HeII}$ observed in quasar absorbers evolve little over 
time from $z=5$ to $z>0$.

\item $\Omega_{\rm HI+HeII}$ observed in quasar absorbers at intermediate and 
high redshift ($z>0$) is high compare with \hi\ inferred from 21-cm
observations of local galaxies.

\item $\Omega_{\rm HI+HeII}$ observed in quasar absorbers over most of the age 
of the Universe (since the Universe was 3 Gyrs old, corresponding to
about $z<2$) is not well constrained.

\end{enumerate}

The first point i) arises mostly from the fact that the most recent
studies make use of modern cosmological parameters (Spergel \e 2003),
including a non-zero $\Lambda$ cosmological constant, which compresses
the high-redshift measurements with respect to the local ones. The
possibility that large number of quasar absorbers are missing in
optically selected quasar surveys is still hotly debated. While radio
surveys looking for DLAs in quasar samples without optical limiting
magnitudes (Ellison \e\ 2001; Ellison \e\ 2004, Prochaska, {\it
private communication}) show that there is not a large number of DLAs
missing, other observational indicators suggest that the DLAs we know
of are not extremely dusty (Pettini \e\ 1997; Ledoux, Petitjean
\& Srianand 2003; Kulkarni \e\ 2005; Murphy \& Liske 2004). It should
be emphasised however that there are two separate issues: i) what is
the dust content of the quasar absorbers we know of today and ii)
which fraction of the quasar absorbers are missed because their
background quasar is not selected in the first place. An extension of
Fall \& Pei (1993) calculations taking into account the most recent
observations addresses both these issues: Vladilo \& P\'eroux (2005)
show that whilst the dust content of the DLAs in current sample is not
too high, the missing fraction is possibly quite important ranging
from 30\% to 50\% at $z=1.8-3.0$.

Concerning points ii) and iii), the number of quasar absorbers observed at
$z>2$ is now reaching hundreds. The mild fluctuations in the redshift
evolution in the range $z=5$ to $z=2$ is within the small error
estimates. Therefore the cosmological evolution of the total gas
mass, $\Omega_{\rm HI+HeII}$, can be approximated to constant in that
redshift range. Yet, radio observations of very large numbers of local
\hi-rich galaxies such as the HIPASS survey (Zwaan \e\ 2005a) indicate a
low value of $\Omega_{\rm HI+HeII}$ at $z=0$. This would imply a fast
evolution of the gas mass between $z\sim0.61$ (Rao, Turnshek \& Nestor
2005) and $z=0$.

The last point illustrates how challenging it still is to find and
study quasar absorbers at intermediate redshifts. Warnings about
$\Omega_{\rm HI+HeII}$ measurements based on small number statistics
have already been issued (P\'eroux \e\ 2004b) and unfortunately future
developments in that direction appear limited given the presently
restricted availability of UV optimised instruments.

\section{Conclusion}

We have presented a new sample of high-redshift sub-DLAs (\nhi $>$
10$^{19}$ cm$^{-2}$) found in the spectra of 17 $z>4$ quasar spectra
observed with the Ultraviolet-Visual Echelle Spectrograph (UVES) on
VLT. The statistical properties of this sample of 21 new sub-DLAs is
analysed in combination with another 10 sub-DLAs from previous ESO
archive studies. This homogeneous sample allows to determine redshift
evolution of the number density of DLAs, sub-DLAs and compare it with
that of LLS taken from the literature. All these systems seem to be
evolving in the redshift range from $z=5$ to $z\sim3$. Assuming that
all the classes of absorbers arise from the same parent population,
estimates of the characteristic radii are provided. $R_*$ increases
with decreasing column density, and decrease with cosmological time
for all systems. The redshift evolution of the column density
distribution, f(N,z), down to \nhi = 10$^{19}$ cm$^{-2}$ is also
presented. A departure from the usually fitted power law is observed
in the sub-DLA regime. f(N,z) is further used to determine the total
\hi\ gas mass in the Universe at z$>$2. The complete sample shows that
sub-DLAs are important at all redshifts from $z=5$ to $z=2$ and that
their contribution to the total gas mass $\Omega_{\rm HI+HeII}$ is
$\sim$20\% or more if compared with the Sloan results (Prochaska \&
Herbert-Fort 2004).  Finally, we discuss the possibility that sub-DLAs
are less affected by the effects of dust obscuration than classical
DLAs.

\section{Acknowledgements}

We would like to thank Nicolas Bouch\'e for letting us publish here
for first time portions of the UVES red spectrum of
SDSS~J0124$+$0044. We are grateful to Sandhya Rao for communicating
results in advance of publication. CP thanks Martin Zwaan for shared
enthusiasm on the topic and the Institute of Astronomy in Cambridge
for a visit during which part of this work was completed. This work
was supported in part by the European Community Research and Training
Network ``The Physics of the Intergalactic Medium''.

\bsp \label{lastpage} 
\end{document}